\documentclass[11pt]{article}
\usepackage{newinutile-2.0}

\usepackage{latexsym,amsmath,amsfonts,amsthm,amssymb,fdsymbol}
\usepackage{epsfig,graphics,color,calc,graphicx,pict2e}
\usepackage{mathrsfs}

\usepackage{subfigure} 

\newlength{\pecettawidth}
\newcommand\blfootnote[1]{%
  \begingroup
  \renewcommand\thefootnote{}\footnote{#1}%
  \addtocounter{footnote}{-1}%
  \endgroup
}

\begin{document}
\title{\normalsize\Large\bfseries 
Learning models for classifying 
Raman spectra of genomic DNA
from tumor subtypes}

\author[1]{Giacomo Lancia}
\author[2]{Claudio Durastanti}
\author[1]{Cristian Spitoni}
\author[2]{Ilaria De Benedictis}
\author[3]{Antonio Sciortino}
\author[2]{Emilio N.M. Cirillo}
\author[4]{Mario Ledda}
\author[4]{Antonella Lisi}
\author[3]{Annalisa Convertino}
\author[3]{Valentina Mussi}


\affil[1]{Mathematical Institute, Utrecht University,
Budapestlaan 6, 3584 CD Utrecht, The~Netherlands.}
\affil[2]{Dipartimento di Scienze di Base e Applicate per l'Ingegneria, 
             Sapienza Universit\`a di Roma, 
             via A.\ Scarpa 16, I--00161, Roma, Italy.}
\affil[3]{Institute for Microelectronics and Microsystems,
          CNR, via del Fosso del Cavaliere, 100, Roma, Italy.}
\affil[4]{Institute of Translational Pharmacology,
          CNR, via del Fosso del Cavaliere, 100, Roma, Italy.}

\date{\empty} 

\maketitle

\begin{abstract}
An early detection of different tumor subtypes 
is crucial for an effective guidance to personalized therapy.
While much 
efforts focus on decoding the sequence of DNA basis 
to detect the genetic mutations related to cancer, it is 
becoming clear that physical properties, including structural 
conformation, stiffness, and shape, as well as biological processes,
such as methylation, can be pivotal to recognize DNA 
modifications. Here we exploit the Surface Enhanced Raman Scattering 
(SERS) platform, based on disordered silver coated--silicon nanowires, 
to investigate genomic DNA from subtypes of melanoma and 
colon cancers and to efficiently discriminate 
tumor and healthy cells, as well as the different tumor subtypes.  
The diagnostic information is obtained by 
performing label--free Raman maps of the dried drops of DNA solutions 
onto the Ag/NWs mat, and leveraging the classification ability of 
learning models to reveal the specific and distinct 
interaction of healthy and tumor DNA molecules with nanowires.
\end{abstract}\blfootnote{Correspondence to 
annalisa.convertino@cnr.it, claudio.durastanti@uniroma1.it, 
and valentina.mussi@cnr.it.}







\section{Introduction}
\label{s:intro} 
\par\noindent
Tumor lineages are divided in multiple subtypes characterized by 
different cell proliferation, migration, and invasion capabilities which, 
on turn, determine cancer aggressiveness and metastatic 
potential \cite{mp2010,pddetal2018}. 
The advent of molecular genetic technology based on DNA sequencing 
methods \cite{scf2009,cz2019} makes it possible to detect DNA mutations 
associated with cancer and related subtypes, but the expensive 
and complex enzyme based target or signal amplification procedures
still prevent genetic analysis to be introduced in the routine diagnostic 
practice \cite{nciUSA}. Thus, there is an urgent need of new strategies 
and tools for an early and accurate detection of the cancer subtypes, 
which will provide an essential guidance for personalized therapeutic 
treatments.

Recent studies have shown that the analysis of physical and mechanical 
properties, such as structural conformation, shape, length, can play 
an important diagnostic role, especially in DNA analysis, to quickly 
recognize changes and alterations associated with diseases and degenerative 
processes \cite{tseytlin2016}. 
In fact, the physical and mechanical characteristics, closely related to 
the chemical structure of DNA, influence the interaction with the 
microenvironment and, ultimately, many cellular processes, such as 
replication, transcription and repair. 

In addition, alterations of DNA methylation have been recognized as 
an important component of cancer development, through hypermethylation 
of tumor suppressor and DNA repair genes, and/or hypomethylation of 
oncogenes. DNA methylation is indeed a biological process by which CH$_3$ 
methyl groups are covalently added to the DNA molecule, mainly at 
cytosine sites within CpG dinucleotides of specific gene promoters. 
In some cancers, such as colon cancer, the detection of hypermethylation, 
may serve as a valuable biomarker for the disease condition \cite{gdbi2021}, 
and that DNA methylome mapping techniques can be exploited for clinical 
diagnostics and personalized treatment 
decisions \cite{Ehrlich2009,Blueprint2016,ke2010}.
Therefore, the development of innovative methods based on the analysis 
and recognition of DNA structural changes can represent an effective 
alternative or complementary approach to the complex genetic methods. 

As we have already demonstrated, disordered nanostructures consisting 
of metal (Ag or Au) coated silicon nanowires (SiNWs) are a powerful 
Surface Enhanced Raman Scattering (SERS) platform, capable both to enhance 
the spectral signal related to specific vibrational modes of  the cellular 
or molecular target, and to provide additional ``probe--signals" associated 
to its physical and chemical interaction with the NWs 
\cite{pcmmb2021,mlcl2021,cmmllbfrl2018,cmm2016}. 
This unique behavior could allow to capture diagnostic information carried 
by structural conformation  and physical charecteristics of healthy 
and malignant DNA molecules, upon Raman mapping of the dehydrated 
aqueous DNA droplets directly deposited onto the nanostructured platform, 
without any  knowledge of the DNA  sequence \cite{mlpmpbllc2021}.
Here, we leverage the sensing capacity of the SERS platform of 
disordered  Ag/SiNWs  with the classification ability of statistical 
models to make a step further and differentiate DNA from tumor subtypes, 
by exploiting their different DNA methylation degree. 

Here, we analyse two different phenotypes of two human 
cancer cell lines and study melanoma and colon cancer progression. 
In particular, 
SK--MEL--28 and A375 cells have been selected as representative 
models for less and more aggressive phenotype of melanoma, 
respectively \cite{nn2017,rct2018}, while CaCo--2 and HT29 as less and more 
methylated colorectal cancer cell models, respectively 
\cite{hkspbmms2012}.
As a control, human keratinocytes cell line (HaCaT), a reliable 
in vitro model for human normal skin, has been considered \cite{bpbhmf1988}.
The analysis of Raman data coming from the drop mapping of the samples 
deposited on the Ag/SiNWs has been performed by using five different 
statistical methods. 

Two of them are essentially geometric and 
involve the full data, that is to say, they 
are based on the 
evaluation of global characteristics of the spectra:
the average and the $\ell^2$ distance from the average.
The remaining three approaches 
are more sophisticated, they take into consideration 
local properties of the spectra, also via an initial 
reduction of the amount of wavenumbers, and are based on the 
PCA analysis, 
the average pooling of the spectra, and the propagation of the spectra 
through a 1--D Convolutional Neural Network (CNN).
We will address to the first two methods as the \emph{global} ones, 
whereas the last three will be called \emph{local} ones.
We will show that all the strategies achieve a very 
high classification accuracy, close to 90\%.
Moreover the local methods allow, also, 
to identify the relevant spectral ranges that appear to be 
decisive for their correct classification. 

In fact, the data distinction process exploits the information 
mainly inferred from two significant spectral regions: a low wavenumber 
(LW region) one, 
between 125 and 550 cm$^{-1}$, containing features related to the 
different physico--chemical interaction of the DNA molecules with the 
nanowires; a second one at higher wavenumbers (HW region), 
between 2300 and 3400 cm$^{-1}$, where vibrational peaks associated 
with the CH/CH$_3$ bonds of the methyl groups are located. Our data 
therefore prove that the remarkable capacity 
of the proposed bioanalytical platform 
to discriminate cancer subtypes
is partly based on its ability to 
highlight the properties associated with the different degree of methylation 
of malignant tumor DNA, and the effects that these epigenetic variations 
have on its mechanical and conformational properties. Given the 
reversibility of the methylation process, the development of novel 
strategies to evicence such mechanism influencing gene expression 
without modification of the DNA sequence, can be extremely interesting, 
not only for diagnostic purposes, but also for therapy approaches.

\section{Materials and Methods}
\label{s:matmet} 
\par\noindent
In this section we describe the 
sample preparation 
procedures, the Raman measurements, and 
the statistical approaches developed to analyse the experimental 
data (see, also, \cite{dcdlslcm2022,mlpmpbllc2021}).

\subsection{Experimental procedures}
\label{s:sperimentale} 
\par\noindent

\subsubsection{Ag/SiNW substrate fabrication}
\label{s:agsinw} 
\par\noindent
Au catalyzed SiNWs have been grown on Si wafers by plasma enhanced 
chemical vapor deposition (PECVD) using SiH$_4$ and H$_2$ as 
precursors at a total pressure of 1~Torr and flow ratio 
SiH$_4$/(H$_2+$SiH$_4$), fixed to $1:10$. The substrate temperature 
during the growth was kept at 350$^\textrm{o}$C, and a 13.6~MHz 
radiofrequency was used to ignite the plasma with power fixed 
at 5~W. A metal coating with a nominal thickness of 90~nm 
was obtained by evaporating an Ag film onto the SiNWs array.

\subsubsection{Cell culture and DNA extraction}
\label{s:cell} 
\par\noindent
As skin cancer model \cite{wlhwgkhoe2016} we used the human melanoma 
cell lines SK--MEL--28 and A375 established
from patient--derived cancer samples and routinely used in skin cancer 
research, which were compared to the human immortalized keratinocyte HaCaT 
as control health skin model. 
As colon cancer model, we used the cell lines CaCo--2 and HT29, which were 
compared to the same human immortalized keratinocyte HaCaT as control 
health skin model. 

All cell lines were cultured in complete Dulbecco's modified Eagle's 
medium (DMEM; Hyclone, South Logan, UT) with high 
glucose (4.5~g$/$L), supplemented with 10\% fetal bovine 
serum (FBS, HyClone), 2~mM--glutamine and 100~IU$/$mL 
penicillin/streptomycin (Invitrogen, Carlsbad, CA). 
Cells were kept at 37$^o$C in a humidified atmosphere 
with 5\% CO$_2$, until medium removal and harvesting by trypsin treatment. 

The cells were passaged every 3--4 days at a 
sub--cultivation ratio of 1:5 and used within 5--20 passages.
The cell pellet resulting from subsequent centrifugation for 5~min 
at 4000~rpm was then processed for genomic DNA extraction. The cells 
were lysed in hypotonic lysis buffer by repeated pipetting, 
incubated 15~min in ice, and centrifuged for 10~min at 2000~rpm 
and 4$^o$C, discarding the supernatant. The extraction of the genomic DNA 
has been performed by incubation for 1~h at 37$^o$C in 
750~$\mu$L of nuclear lysis buffer, followed by a treatment 
with 250~$\mu$L of NaCl 6M and final centrifugation for 15~min 
at 2000~rpm and 4$^o$C. The supernatant containing genomic DNA was 
recovered and then precipitated adding EtOH 100\%. 
The DNA pellet obtained by further centrifugation for 10~min 
at 2000~rpm and 4$^o$C was washed in EtOH 70\%, 
centrifuged again for 10~min at 7500~rpm and 4$^o$C 
and re--suspended in DNase free H$_2$O to obtain a 20~ng$/\mu$L solution. 
The final samples are prepared by depositing one drop of 
the DNA solution on Ag/SiNWs substrates
coming from the very same batch.

\subsubsection{Raman analysis}
\label{s:raman} 
\par\noindent
A DXR2xi Thermo Fisher Scientific Raman Imaging Microscope has been 
used to collect a Raman map of all the DNA drops deposited onto 
the nanostructured substrates and left to dry in air at Room Temperature. 
The maps have been collected by using a 532~nm laser source, with 
a 1~mW excitation power and a 50x objective in a backscattering 
configuration. 

Each spectrum composing the map resulted from 4 accumulations 
lasting 5~ms each. The map step size has been fixed in 
4~$\mu$m so as to obtain about 4000 independent spectra for each drop. 
For each sample, HaCaT, SK--MEL--28, A375, CaCo--2, and HT29, 
the entire measurement process has been repeated 5 to 10 times. 

\subsection{Data set and pre--processing}
\label{s:dataset} 
\par\noindent
We performed a pairwise comparison of the experimental data by 
comparing each time two sets of Raman spectra coming from two 
different samples. More precisely, we considered six cases: 
1) HaCaT vs.~A375, 
2) HaCaT vs.~SK--MEL--28,
3) SK--MEL--28 vs.~A375,
4) HaCaT vs.~CaCo--2, 
5) HaCaT vs.~HT29, 
and 
6) CaCo--2 vs.~HT29. 
For each case the two involved samples are conventionally called 
\emph{first} and \emph{second} sample. 

In each comparison the data set consisted of 
$N=4000$ spectra, $2000$ for each sample. 
The spectra have 
been randomly chosen in the central part of the 
droplets to exploit maximally 
the interaction between the DNA
molecules and the nanostructured substrate.
Since the spectra are collected 
at points of the droplet at distance larger than, or equal to, 
$4~\mu$m we assume that the data are independent \cite{dcdlslcm2022}. 

Large part of the collected spectra share similar features.
The few which are substantially different from the others have 
been considered
outliers and removed from the analysis. To do this, 
we built a decision surface by adding and subtracting three times 
the (point--wise) empirical standard deviations to the average spectra
and discarding those spectra 
featuring at least one point outside the decision surface.
Moreover, we smoothed the data by filtering 
the original raw spectra with 
the Savitzky--Golay algorithm \cite{SG64} (see also \cite{ZK13})
over a window of $90$ data points treated as convolution coefficients.

\subsection{Statistical approaches}
\label{s:stat} 
\noindent 
We propose five different models to classify the spectra. 
Two are rather simple, essentially geometric, and based on the 
evaluation of global characteristics of the spectra: 
the average and the $\ell^2$ distance from the average.
The other three are more sophisticated, they take into consideration 
the local properties of the spectra, and are based on the 
PCA analysis, 
the average pooling of the spectra, and the propagation of the spectra 
through a 1--D Convolutional Neural Network (CNN).

The spectra are split into training and test sets, 
the first used to tune the parameters of the models here proposed, 
the second to validate them.   
We deal with a binary target variable $W$, whose outcomes 1 and 0 correspond 
to the DNA molecules of the first and the second sample, 
respectively.

\subsubsection{Logistic regression on global average (LRA)}
\label{s:ave} 
\par\noindent
The first model we proposed is based on a very simple idea, i.e., the global mean of the spectral intensity is exploited as the unique predictor of a logistic regression model.
Thus, the mean value provides a representation of the spectra 
in both the LW and HW regions.
This method does not involve highly detailed data analysis and is meant to provide a basic tool to classify Raman Spectra. In particular, it is of interest to check whereas such a simple approach can be sufficient to identify properly tumoral DNA.
The model is penalized by means of an $L^2$ regularization with setting 1 as shrinkage parameter. 
The model is validated by means of a ten--fold cross--validation, and its goodness is assessed by the average score of the ROC--AUC 
(Receiver Operating Characteristic -- Area Under the Curve) curve over the 
ten cross--validation folds. 

\subsubsection{Evaluation of $\ell^2$ distance (L2D)}
\label{s:geo} 
\par\noindent
The second method here proposed is based on the analysis of some 
geometric features of the whole considered portion of the spectra and has 
already been presented in \cite{dcdlslcm2022} and it can be described as follows. The training set is used to compute 
average spectra of the two samples, represented by the 
column vectors of $\mathbb{R}^p$, 
$h^1$ for the first sample and 
$h^2$ for the second one. Then, the $\ell^2$ distance between both these averages and each spectrum belonging to the test set and represented by the 
$i$--th row of the data matrix $\mathbf{X}$ is computed as follows,
\begin{equation}
	\label{geo000}
	\textup{d}^k(i)
	=
	\sum_{s=1}^p|x_{is}-h^k_s|^2
	\;\;\;\textup{ for }
        k=1,2
	.\end{equation}
A classifier for each spectrum is then built thanks to the following outcome binary function
\begin{equation}
	\label{geo020}
	g_\textup{out}(i)
	=
	\mathbb{I}\{\tau \textup{d}^1(i)\le(1-\tau)\textup{d}^2(i)\}
	,
\end{equation}
where $\tau\in[0,1]$ is an optimization parameter, again 
optimized as above with a ten--fold validation.
The outcome is set as equal to $1$ if the spectrum is identified as 
coming from the first sample.

\subsubsection{Logistic regression on average pooling (LRP)}
\label{s:reg} 
\par\noindent
The third method here proposed is also based on a logistic regression model, where the input features are computed by applying the average pooling operator on the input spectra.
Thus, the spectral domain of each spectrum is divided into non--overlapped and non--equispaced sub--domains and for each subset the mean value is computed.
Such an approach aims to pre--process and represent the profile of each spectrum with a lower number of explanatory variables. 
Depending on the binary task to solve, a finer or coarser partitioning of the spectral domain can be chosen; here a four and three--feature representation has been set for the LW and HW regions, respectively.

Again, the logistic regression model is penalized by the $L^2$ regularization with the shrinkage parameter equal to 1 and the model validation follows a 
ten--fold cross--validation procedure.

The \emph{permutation importance} technique \cite{breiman2001random} is here used to investigate which input features mainly support the predictions of the logistic regression model.
This technique is often employed to inspect if the random shuffling of one column feature can drastically decrease the accuracy of the model.
Indeed, a random permutation in a column feature causes the break of the correlations between that specific explanatory variable and the target variables and produces a drop in the predictive performance of the model.
To quantify the degradation of the accuracy due to the permutations on one column feature, we evaluate the difference between the ROC--AUC of the model and the average ROC--AUC after applying a finite number of permutations.
Such a quantity is labeled as the \emph{importance score} of 
one column feature. 
In this case, 30 distinct permutations on each column feature have been applied.

\subsubsection{Logistic regression on PCA components (PCA)}
\label{s:pca} 
\par\noindent
We use the notation 
borrowed from \cite[Chapter~1 and Paragraph~8.2.1]{htw2015}
and already used in \cite{dcdlslcm2022}.
Any single spectrum, in the considered interval,
is represented as a column vector $x\in\mathbb{R}^p$.
By collecting the $N$ row vectors $x^\dag$, where $\dag$ 
denotes transpositions, we construct the 
$N\times p$ matrix $\mathbf{X}$ which represents the entire data set
in the considered interval. 
The $j$--th column of $\mathbf{X}$ is 
the collection of the $N$ observations of the $j$--th variable, 
namely, the intensity corresponding to the $j$--th value of the Raman shift. 

Thus, we compute the $N\times p$ matrix $\mathbf{Y}$ 
by centering $\mathbf{X}$ with respect to the 
columns (i.e., the Raman shift). 
A \emph{principal components analysis} is then obtained by the  
eigendecomposition of the  empirical covariance matrix 
$\textbf{Y}^\dag\textbf{Y}$ \cite{Jackson91}. 
The \emph{principal components directions}
$v_1,\dots,v_p\in\mathbb{R}^p$ are computed and
we call $i$--th \emph{principal component loadings}
the $p$ elements 
of the column vector $v_i$.
The projection $\mathbf{z}_i=\mathbf{Y}v_i\in\mathbb{R}^N$ is called $i$--th 
\emph{principal component} (PC) of the data $\mathbf{Y}$
and its variance of each PC is given by the corresponding 
eigenvalue. The variance concentrates on the 
first $m$ principal components, allowing us to neglect in 
the next step all the other $p-m$ components. This is way we call 
``local" this method. 
 
The selected first $m$ principal components $z_1,\dots,z_m$ are 
thus interpolated to build a logistic regression model to estimate 
the  probability mass function of the binary 
target variable $W$ by 
\begin{equation}
	\label{pca020}
\textup{Pr}(W=1|z_1,\dots,z_m)=\frac{e^{\beta_0+\sum_{i=1}^m\beta_iz_i}}
                                    {1+e^{\beta_0+\sum_{i=1}^m\beta_iz_i}}
\;\;\textup{  and }\;\;
\textup{Pr}(W=0|z_1,\dots,z_m)=\frac{1}{1+e^{\beta_0+\sum_{i=1}^m\beta_iz_i}},
\end{equation}
where $\beta_i\in\mathbb{R}$ for $i=0,\dots,m$.
In this case, we consider an optimization 
parameter $\lambda\in[0,1]$ such that we associate 
the outcome for the binary variable $W=1$ to each set 
of components $z_1,\dots,z_m$ if  
$\textup{Pr}(W=1|z_1,\dots,z_m)\ge\lambda$ and $W=0$ otherwise. 
The tuning parameter $\lambda$ is estimated
by means of a ten--fold cross--validation procedure
(see, \cite[Ch.~7]{HTF09}). 

The original sample is randomly partitioned into ten equally sized subsamples. A subsample is kept as test set, while the other 
nine ones are used  as training data. Then the accuracy of both the methods are evaluated,  while the cross--validation process is repeated ten times, paying attention to use each round a different subsample as test group. The ten results are then averaged to compute a single estimation. 
The advantage of this validation strategy is that all observations are
used at the same time for both training
and testing and each observation is used for testing exactly once. 

\subsubsection{1--D Convolutional Neural Network (CNN)}
\label{s:cnn} 
\par\noindent
1--D CNN represents a type of feed--forward neural network designed to solve a broad class of classification tasks when the input features are precisely 1--D grid--structured data\cite{yoon2014convolutional, mesuga2022lepton, di2022prediction, cui2016multi}.
Such a class of models combines convolutional and max--pooling operators to encode the sequentiality of the patterns contained in the input data.
As a result, the optimization of the weights defining the convolutional filters of the convolutional layers aims to give the most linearized latent representation of the input Time--Series.  
In our case, therefore, we regarded the spectra as some Time--Series 
whose ``temporal evolution'' takes place along the spectral domain.

Before being propagated thought the layers of the CNN model, the spectra are neither rescaled nor transformed furtherly.
The model, therefore, is validated by means of ten--fold cross--validation.
To assess the goodness of the model, we compute the ROC--AUC score on each fold; the average AUC score over all the cross--validation folds is to be intended as the goodness of the model. 
The Standard Error Mean is used to estimate the error on the average ROC--AUC score. 

The design of our CNN is purely convolutional, i.e. it consists of a sequential combination of \emph{Convolutional Layers} followed by \emph{Max--Pooling Layers}. 
The non--linear activation function is embedded in the convolutional layer; in specific, we opted for a \emph{softplus} function, i.e. $\phi(x) = \log{(1+\exp{(x)})}$. 
Dropout layers \cite{srivastava2014dropout} with dropout rate of 0.25 are also employed to contrast overfitting.
Each convolutional layer possesses 64 filters whose convolutional masks have an amplitude of 3 pixels; the pooling size of the pooling layers is equal to 2.
The sequence of convolutional and pooling layers is then repeated three times; the resulting feature map is therefore flattened by means of a \emph{Flatten Layer}.
Finally, the flattened feature map is propagated through one \emph{Fully--Connected Layer} with 16 output nodes and softplus activation function.
This particular layer returns the latent representation of the spectra.
The latent representation is therefore propagated through a \emph{Fully--Connected Layer} with a sigmoid activation function and one output node, that is the output node of the CNN.
During the training phase, the ADAM \cite{kingma2014adam} algorithm is used to optimize the Binary--Cross Entropy loss function. 
The batch size and learning rate are set equal to 64 and 0.001, respectively.

As we know, feed--forward neural networks are usually regarded as black--box models whose activity cannot be expressed by means of closed forms.
However, we can visualize the impact that a single input feature has on the final predictions by means of the Vanilla Gradient method \cite{simonyan2013deep}.
Such a method allows the construction of \emph{saliency maps}
based on the evaluation of the gradients $\frac{\partial o}{\partial X_i}$, with $o$ the output value of the CNN and $X_i$ the $i$--th input feature.
The evaluation of the score--derivative $\frac{\partial o}{\partial X_i}$ is nothing but the change needed at the pixel $X_i$  to affect the class score the most \cite{simonyan2013deep}.
Note that, in our case, $X_i$ is exactly the intensity of the $i$--th Raman shift.
The visualization and the interpretation of saliency maps, however, can often suffer from scale problems.
Such an inconvenience is often adjusted by applying a monotone function to map the values of the score derivative into the desired interval. 
We, therefore, opted for the empirical cumulative functions of the score--derivatives themselves, i.e. we construct a saliency map that is specific to the test set.
Hence, we feed the CNN model with the instances of the test set; next, we backpropagate the output scores via the Vanilla Gradient method and finally apply the desired empirical distribution function on the score derivatives.

\begin{figure*}[t]
	\centering
	\includegraphics[width=.55\columnwidth]{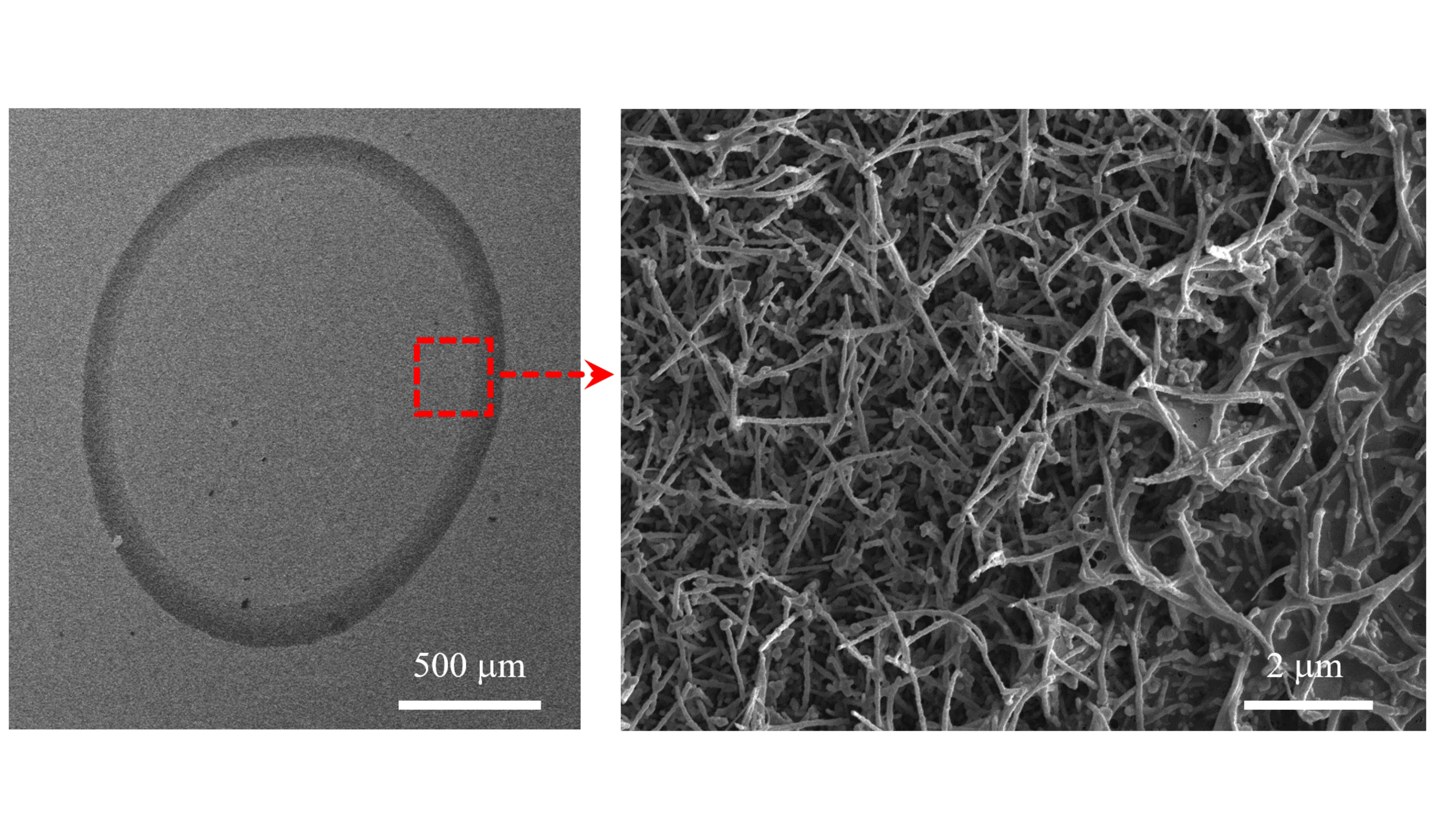}
	\caption{SEM images of a representative DNA drop on Ag/SiNW 
after water evaporation (left panel) and high magnification image of the 
area inside the red 
square
(right panel).}
	\label{f:fig00}
\end{figure*}

\begin{figure*}[t]
	\begin{picture}(80,180)(-10,0)
        \put(0,0){
	\includegraphics[width=.45\columnwidth]{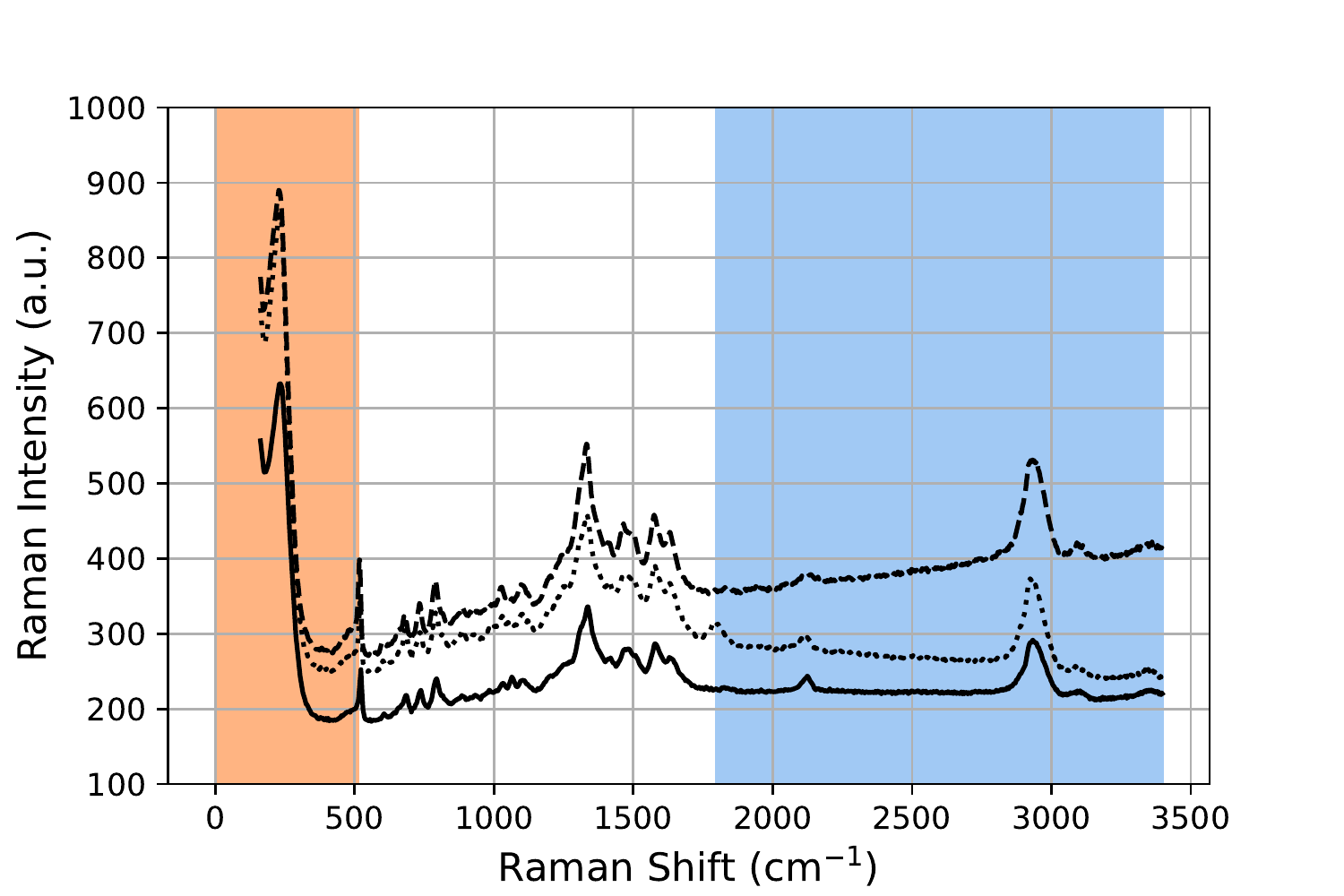}
        }
        \put(260,0){
	\includegraphics[width=.45\columnwidth]{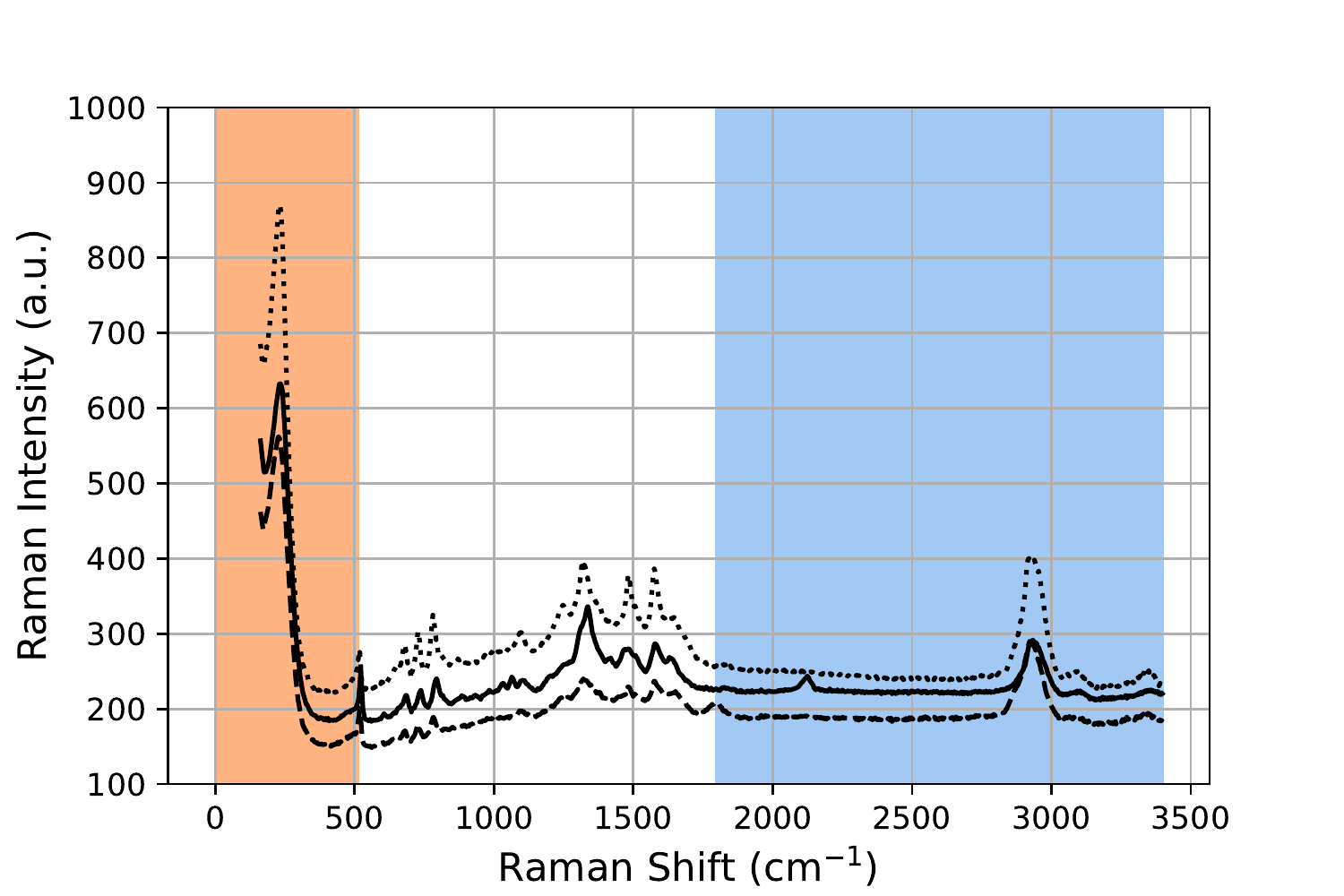}
        }
	\end{picture}
	\caption{Left: average Raman spectrum for the
         HaCaT (solid), A375 (dotted), SK--MEL--28 (dashed) samples.
	 Right: average Raman spectrum for the
         HaCaT (solid), CaCo--2 (dotted), HT29 (dashed) samples.
                }
	\label{f:fig01}
\end{figure*}

\section{Results and discussion}
\label{s:risultati} 
\par\noindent 
Drop casting has been used to deposit healthy and cancer DNA 
solutions on the Ag/SiNWs platform. Figure~\ref{f:fig00} (left panel) 
shows a representative dried DNA drop characterized by the typical 
coffee--ring pattern. The magnified SEM image of the area in the red 
square,
reported in the right panel, shows the morphology of 
the disordered mat of Ag/SiNWs, which are 2--3~$\mu$m long and have 
diameters ranging from 80 to 150~nm. Furthermore, it is possible 
to observe a slight sticking of the wires due to the presence of 
adsorbed DNA.  Raman maps were collected in the central part of the 
drops, to exploit maximally the interaction between the DNA molecules 
and the nanostructured substrate. Figure~\ref{f:fig01}, left and right panels,  
report the average Raman spectra calculated over the entire maps 
for HaCaT with the two melanoma phenotypes, and HaCaT with the two 
colon cancer phenotypes, respectively. Some specific features can be 
recognized: i) the bands directly ascribed to the DNA molecules, i.e., 
those located between 600 and 1200~cm$^{-1}$ due to aromatic in--plane 
bending vibrations of the bases and stretching vibrations of the 
phosphate moiety;  ii) the peak at about 234~cm$^{-1}$, associated 
with the metal--nitrogen (Ag--N) stretching vibration mode of the 
generated surface bond between the deposited nucleotides and the Ag 
coverage of the NWs \cite{gm2002,sm1986,j2002,tmrr2015}; iii) the peak 
at about 514~cm$^{-1}$, originated from the SiNWs themselves, that 
produce a detectable Si signal even through the Ag coating; iv) the 
pronounced large band at about 2934~cm$^{-1}$, corresponding to the 
stretching vibrations of the CH$_2$ and CH$_3$ groups \cite{cz2019}.

The aforementioned Raman features are thus related not only to 
intrinsic chemical characteristics of the DNA molecules, but 
also to their physical properties, which influence the molecule 
arrangement on, and interaction with, the NWs, carrying important 
diagnostic information. In fact, the band around 234 cm$^{-1}$ takes 
into account the specific DNA molecule adsorption on the nanostructured 
Ag surface through the N atoms of the basis rings, so that unrepaired 
oxidative DNA damage, or a different stiffness, can influence the Raman 
signal at that band. The Si peak at 514 cm$^{-1}$ coming from the 
substrate provides information on the surface distribution of the 
molecules: a different DNA conformation results in a diverse substrate 
coverage and causes a consequent variation of the peak intensity. Finally, 
the band at 2934 cm$^{-1}$,  comprising contributions coming from C--H 
vibrations, is clearly conditioned by the degree of DNA methylation.  
On the basis of these considerations, we performed our statistical 
analysis by focussing our attention on two principal spectral ranges: 
the \emph{low wavenumbers} (LW) 
region consisting of $p=221$ spectral points 
with wavenumber ranging from 125.25~cm$^{-1}$ to 549.27~cm$^{-1}$
(orange selection in figure~\ref{f:fig01})
and
the \emph{high wavenumbers} (HW) region consisting of $p=570$ spectral points 
with wavenumbers from 2303.16~cm$^{-1}$ to 3399.83~cm$^{-1}$
(light blue selection in figure~\ref{f:fig01}).

We first discuss our results for the melanoma data, to concentrate
subsequently on the colon tumor ones. 
Here, our aim is to show that the applied classification
methods achieve a very good 
accuracy when applied to different experimetal settings and targets.

\begin{table}[!htb]
\begin{center}
\begin{tabular}{l|ccccc}
 & LRA & L2D & LRP & PCA & CNN \\
\hline
HaCaT vs.~A375 LW & 0.99 & 0.82 & 0.99 & 0.99 & 0.99 \\
HaCaT vs.~A375 HW & 0.97 & 0.90 & 0.96 & 0.99 & 0.96 \\
\hline
HaCaT vs.~SK--MEL--28 LW & 0.99 & 0.89 & 0.99 & 1.00 & 0.99 \\
HaCaT vs.~SK--MEL--28 HW & 0.99 & 0.99 & 0.99 & 0.99 & 0.99 \\
\hline
SK--MEL--28 vs.~A375 LW & 0.61 & 0.62 & 0.88 & 0.93 & 0.90 \\
SK--MEL--28 vs.~A375 HW & 0.95 & 0.94 & 0.99 & 0.99 & 0.99 \\
\hline
\end{tabular}
\end{center}
\caption{AUC values for the five methodes proposed 
in Section~\ref{s:stat} for the three melanoma related 
cases listed 
at the beginning of Section~\ref{s:dataset} in the
low wavenumber (LW) and high wavenumber (HW) spectral regions.
The first column specifies the case and the spectral region. 
In the following five columns the AUC values are reported 
for the methods logistic regression on global average (LRA), 
computation of $\ell^2$ distance (L2D), 
logistic regression on average pooling (LRP), 
logistic regression on PCA components (PCA),
1--D convolutional neural network (CNN).
}
\label{t:auc}
\end{table}

\begin{figure*}[!htb]
	\begin{picture}(80,140)(-10,0)
        \put(0,0){
	\includegraphics[width=.4\columnwidth]{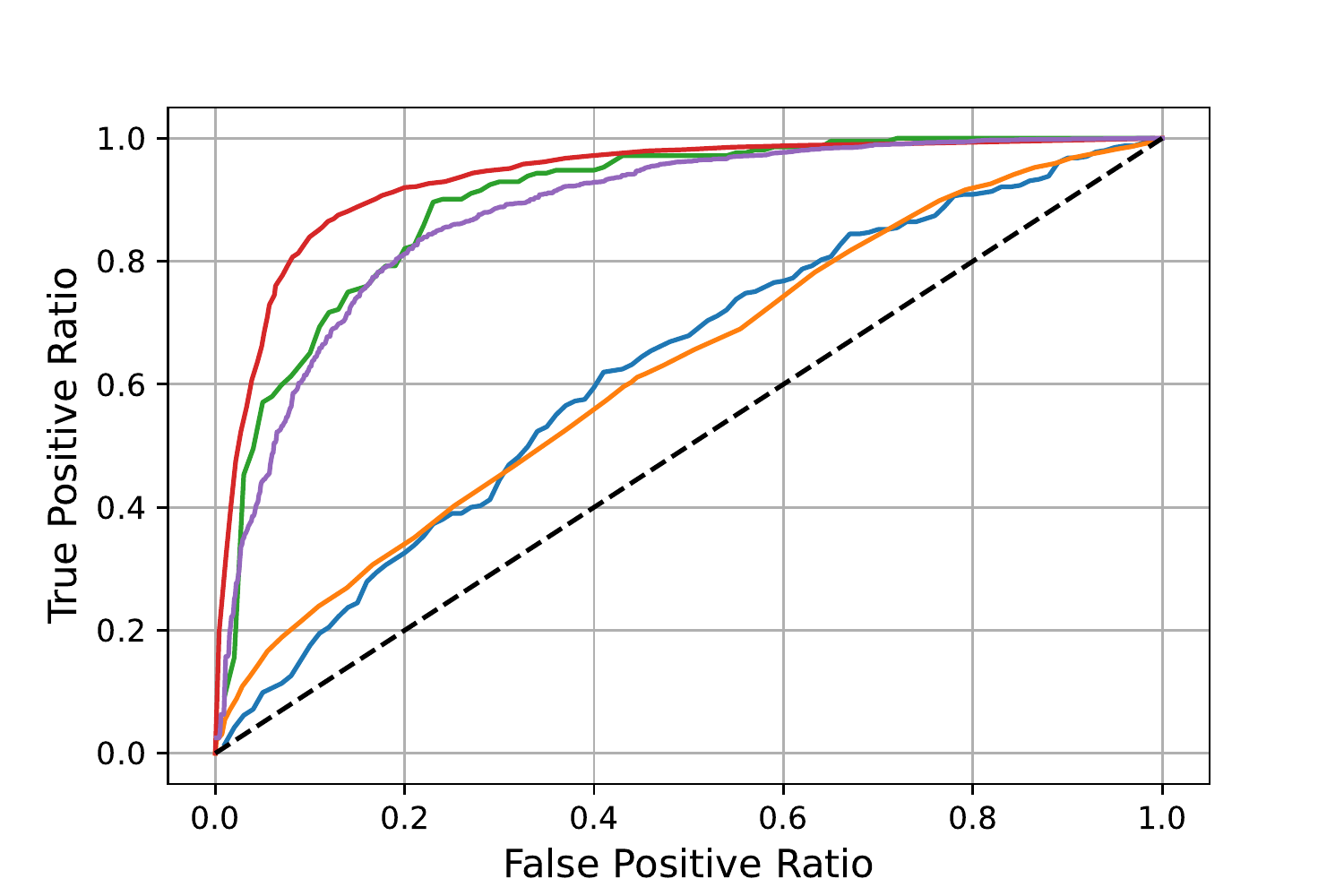}
        }
        \put(260,0){
	\includegraphics[width=.4\columnwidth]{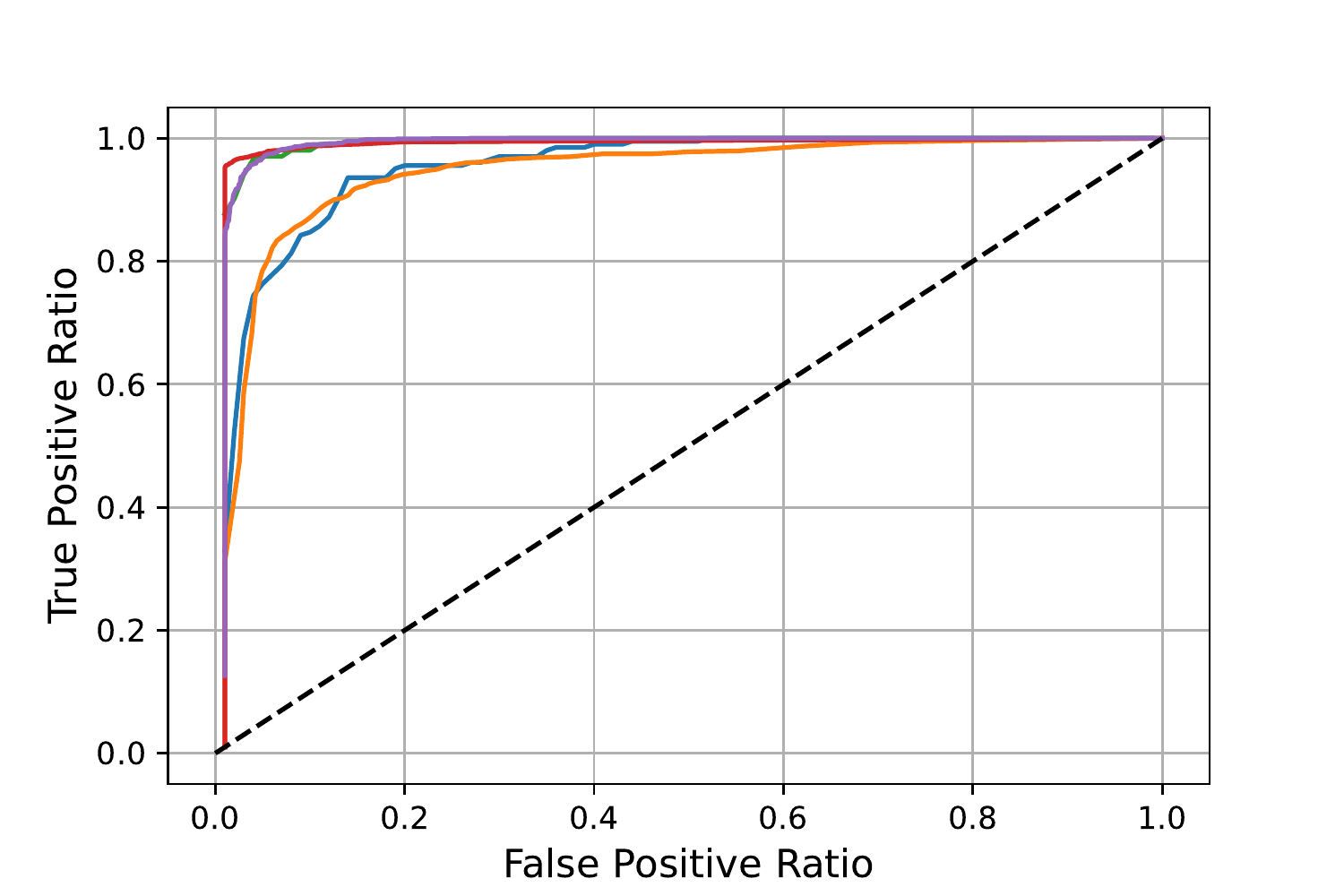}
        }
	\end{picture}
	\caption{SK--MEL--28 vs.~A375 comparison.
ROC graphs for the methods 
logistic regression on global average (blue), 
computation of $\ell^2$ distance (orange), 
logistic regression on average pooling (green), 
logistic regression on PCA components (red),
1--D convolutional neural network (purple)
in the low (left) and high (right)
wavenumber regions.}
\label{f:fig02}
\end{figure*}

\begin{figure*}[!htb]
	\begin{picture}(80,150)(-10,0)
        \put(0,0){
	\includegraphics[width=.4\columnwidth]{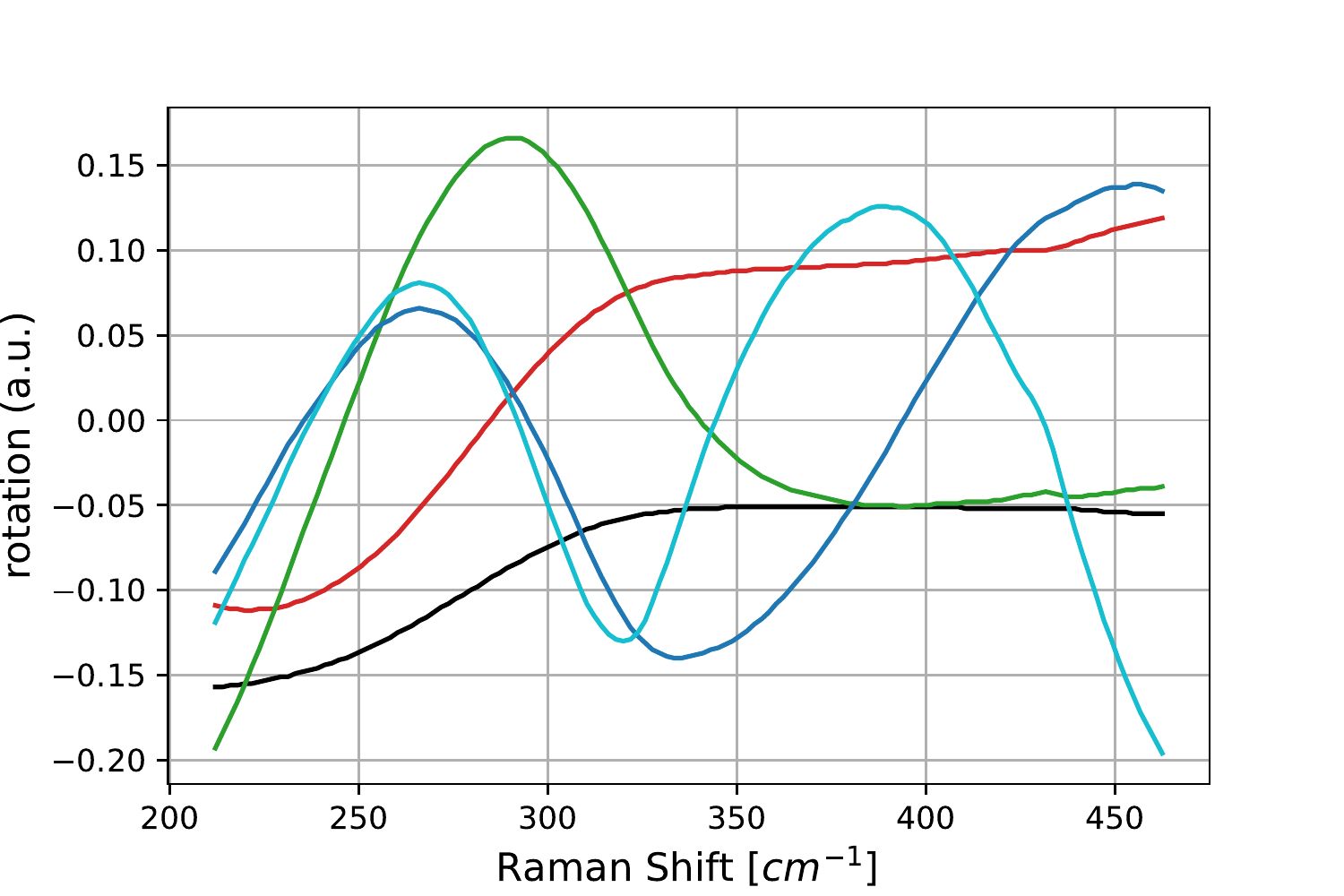}
        }
        \put(240,0){
	\includegraphics[width=.4\columnwidth]{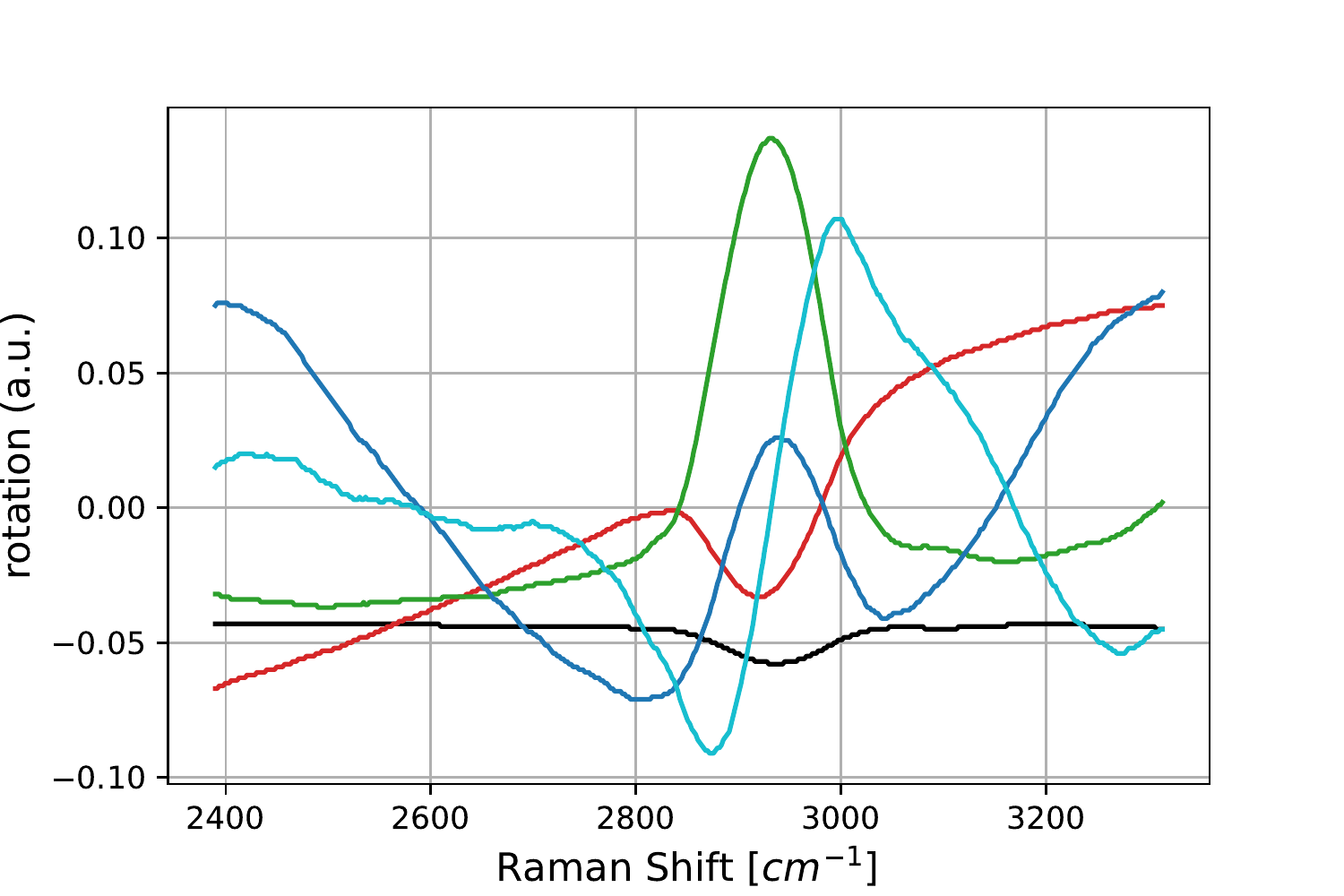}
        }
	\end{picture}
	\caption{SK--MEL--28 vs.~A375 comparison.
Loadings of the first five 
PCA components 
in the LW (left) and HW (right)
wavenumber regions.
Black, red, green, blue, and light blue, respectively, 
for components from one to five.  
}
\label{f:fig03}
\end{figure*}

\begin{table}[!htb]
\begin{center}
	\begin{tabular}{ c | c c c c c }
 LW region & PC1 &      PC2   &   PC3 &     PC4 &     PC5  \\  
 \hline 
Standard deviation  &   7684.2752 & 251.63904 &136.90699 & 51.71480 & 36.92418 \\
Proportion of Variance &   0.9984  & 0.00107 &  0.00032 & 0.00005 &  0.00002   \\
Cumulative Proportion &    0.9984 &   0.99948 &  0.99980 & 0.99984 &  0.99987  \\
\hline
 HW region & PC1 &      PC2   &   PC3 &     PC4 &     PC5  \\
 \hline   
Standard deviation  &   5128.2426 &190.61525 &53.14363 &28.83052& 19.76634 \\ 
Proportion of Variance &   0.9984 &  0.00138 & 0.00011&  0.00003 & 0.00001\\  
Cumulative Proportion   &  0.9984  & 0.99983 & 0.99994 & 0.99997&  0.99998 \\
\hline 
\end{tabular}
\end{center}
\caption{SK--MEL--28 vs.~A375 comparison.
Standard deviation, proportion and cumulative proportion of variance related to the first five PCA components in LW and HW regions, respectively.
}
\label{t:propvar}
\end{table}

\subsection{Analysis of melanoma data}
\label{s:melanoma} 
\par\noindent 
In order to evaluate the ability of the methods proposed in 
Section~\ref{s:stat} to classify the Raman spectra we first 
report in Table~\ref{t:auc} the AUC values obtained with the 
different methods for the HW and LW regions of the 
spectra.
The associated ROC graphs are reported in 
figure~\ref{f:fig02}.

We first note that the low wavenumber part of the spectra 
allows to distinguish healthy (HaCaT) and tumor (SK--MEL--28 or A375) 
samples with all the proposed methods. The less performing is the 
one based on the computation of the $\ell^2$ norm.
On the other hand, this spectral region does not seem 
to ensure a good classification of tumor sybtypes, indeed, 
in the comparison between the SK--MEL--28 and the A375 data only 
the PCA method is able to perform with an AUC larger than 0.9.
The other two local methods work reasonably well with a 0.88 AUC,
while the global methods performance is absolutely poor. 

The tumor samples are, on the contrary,
very well classified and distinguished with all the proposed 
methods applied to the high wavenumber part of the spectra. 

The performance of the local methods is generally better than that of 
the global ones, but the reason why they are particularly useful 
is that they provide us with a detailed information about 
the physical and the chemical phenomena which allow the 
classification of the tumor subtypes. 

Thus, we restrict our discussion to the SK--MEL--28 vs.~A375 
comparison and show how our statistical analysis provides
information at the physical and chemical level. 

\begin{figure*}[!htb]
	\begin{picture}(80,140)(-10,0)
        \put(0,0){
	\includegraphics[width=.4\columnwidth]{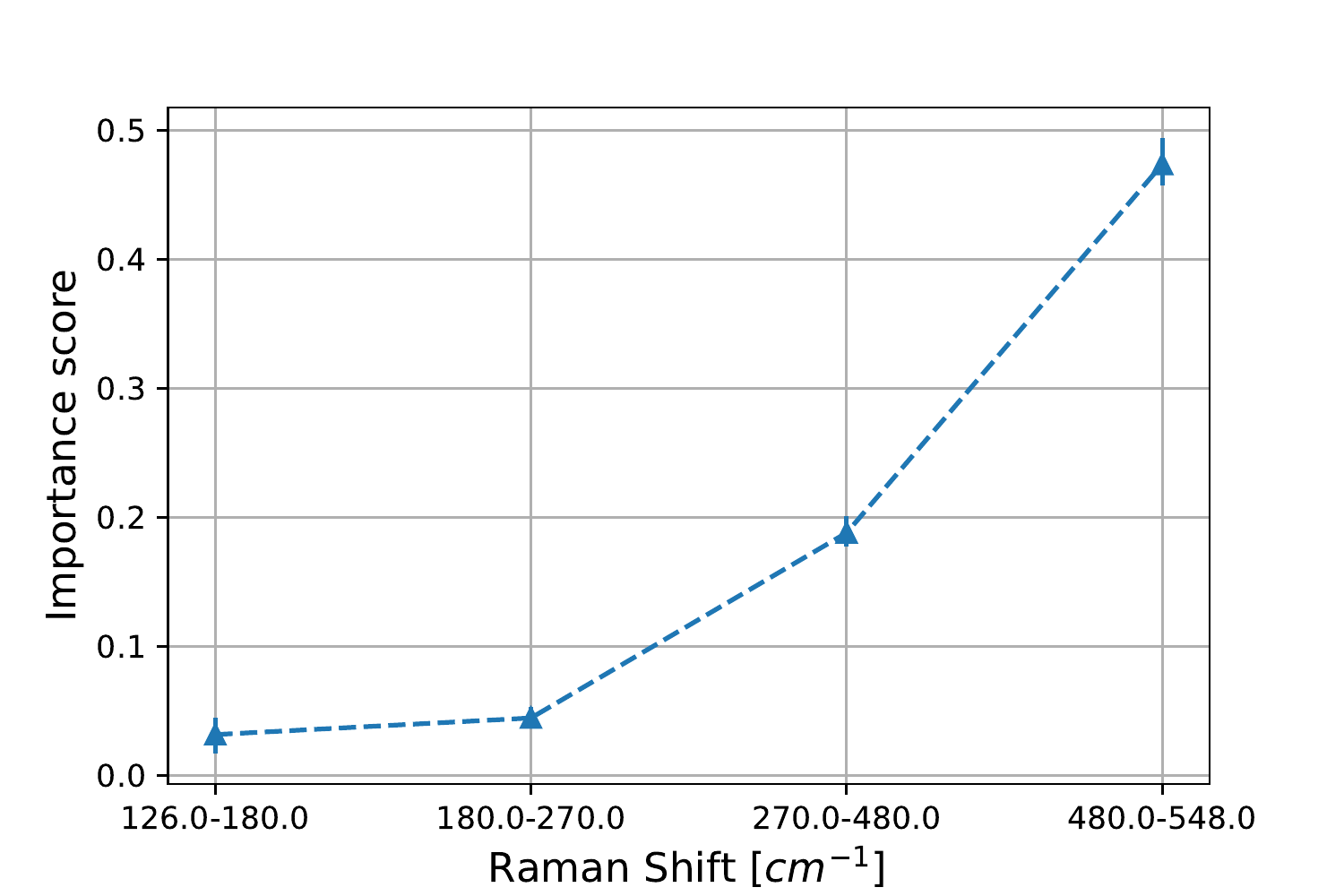}
        }
        \put(240,0){
	\includegraphics[width=.4\columnwidth]{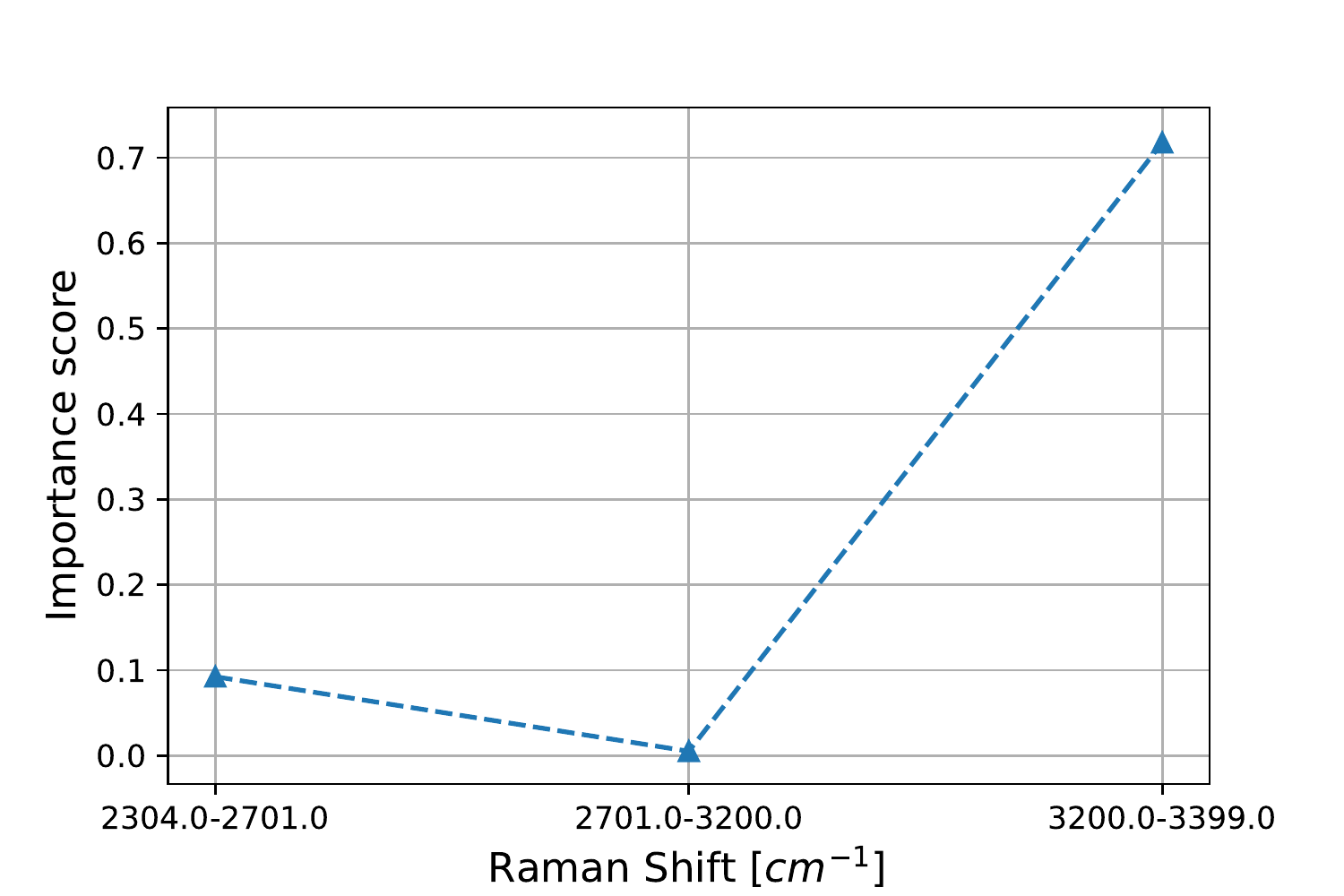}
        }
	\end{picture}
	\caption{SK--MEL--28 vs.~A375 comparison.
Importance scores 
(see Section~\ref{s:reg})
for the logistic regression on 
average pooling for LW (left) and HW (right) regions. 
On the x--axis the sub--regions and on the y--axis the average 
importance scores.
The error bars represent the 95\% confidence interval.
}
\label{f:fig04}
\end{figure*}

\begin{figure*}[!htb]
	\begin{picture}(80,140)(-10,0)
        \put(0,0){
	\includegraphics[width=.4\columnwidth]{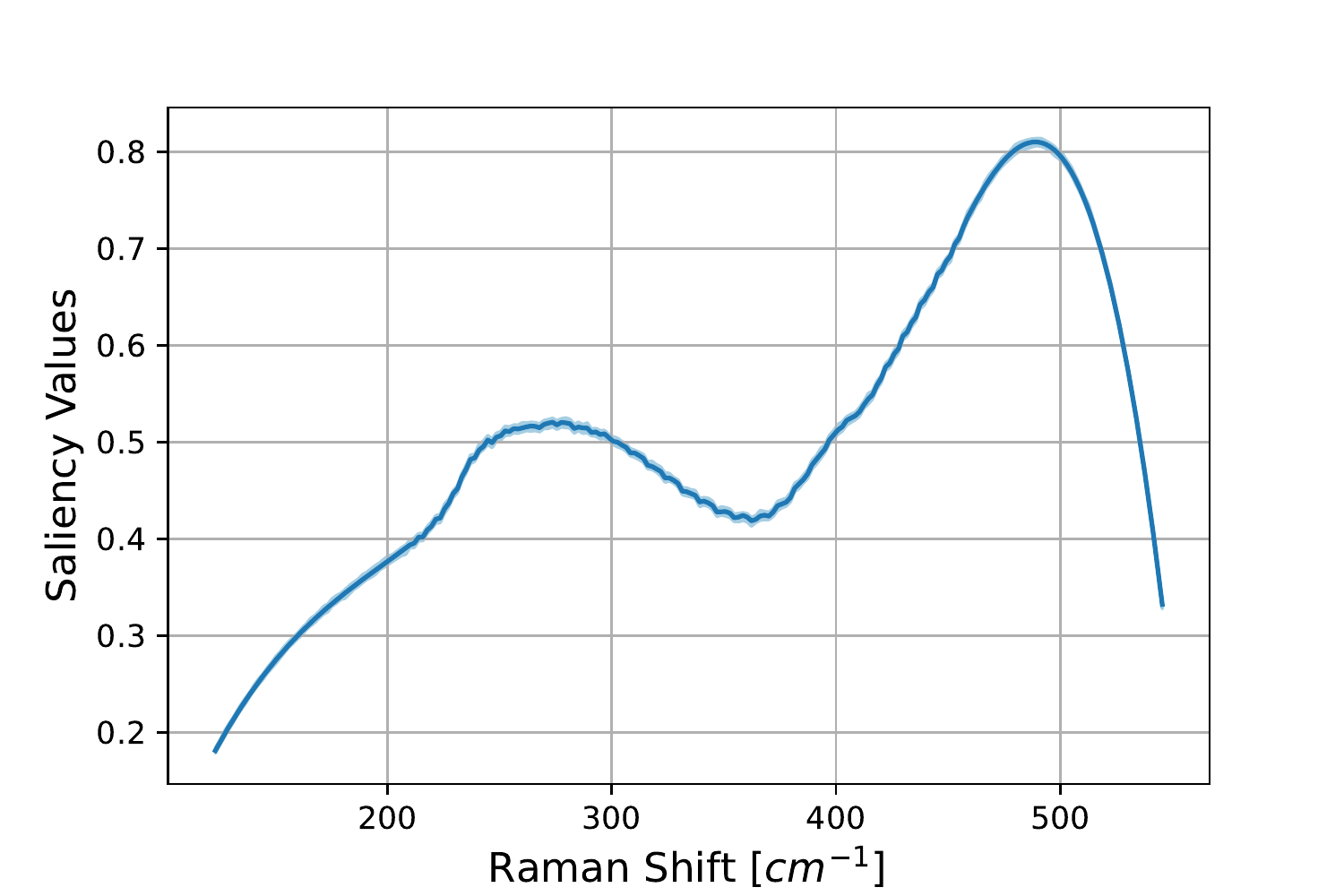}
        }
        \put(240,0){
	\includegraphics[width=.4\columnwidth]{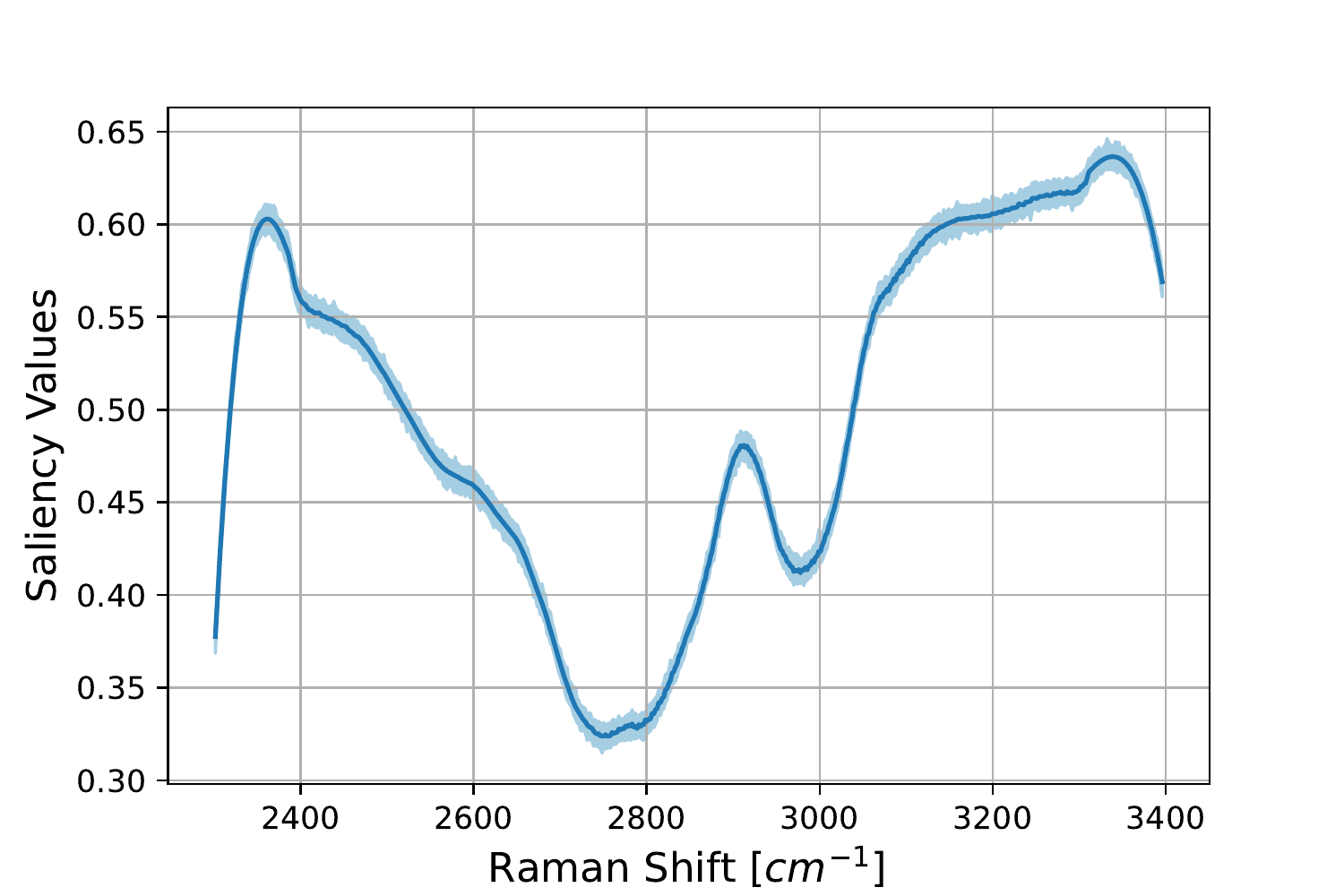}
        }
	\end{picture}
	\caption{SK--MEL--28 vs.~A375 comparison.
Saliency maps via Vanilla Gradient algorithm 
(see Section~\ref{s:cnn})
of 
the 1D--CNN models for both LW (left) and HW (right) regions. 
On the x--axis the spectral domain while on the y--axis the 
average saliency values. 
The light blue and the light red areas represent the 95\% confidence 
interval.}
\label{f:fig05}
\end{figure*}

\begin{figure*}[!htb]
	\begin{picture}(80,140)(-10,0)
        \put(0,0){
	\includegraphics[width=.4\columnwidth]{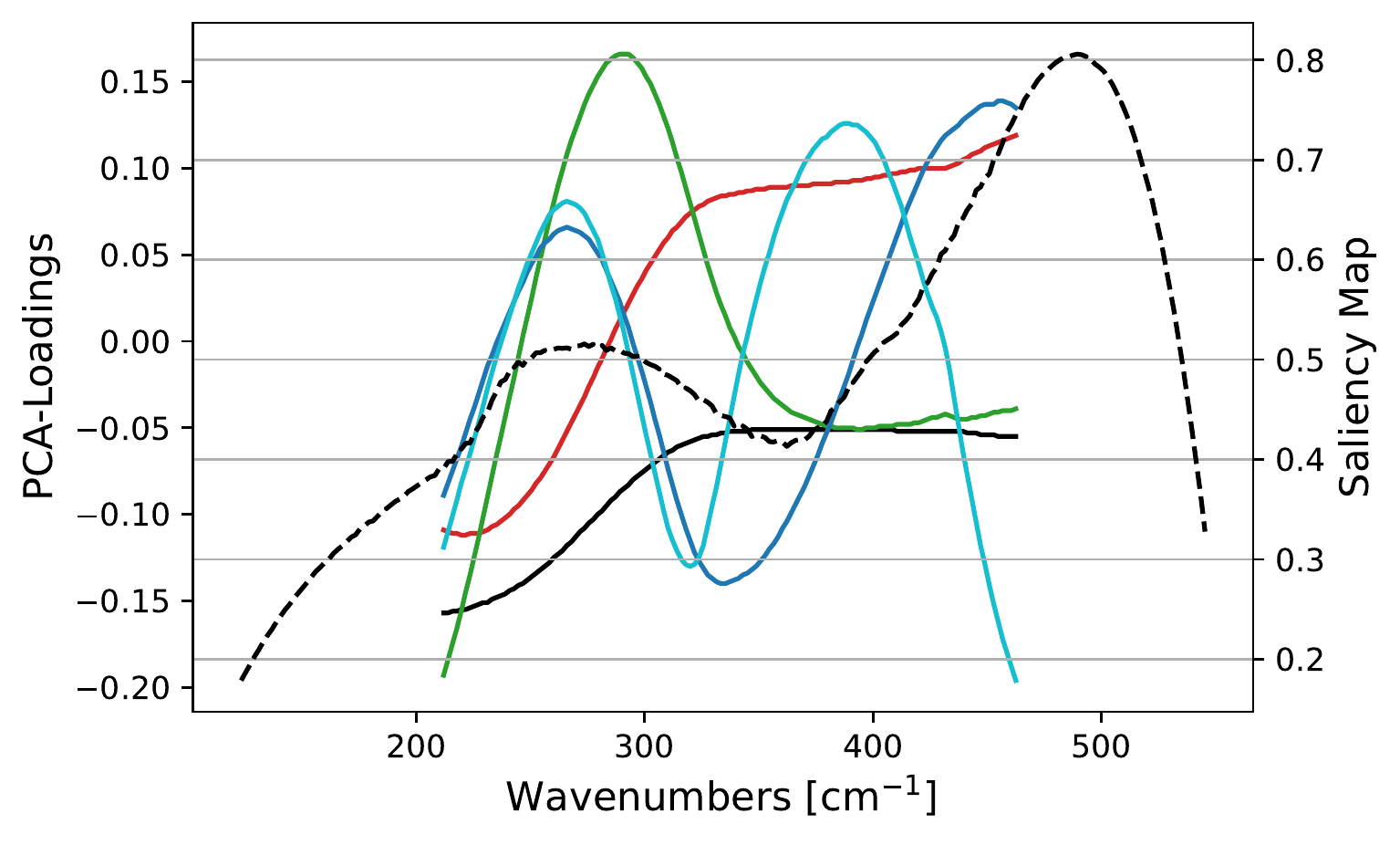}
        }
        \put(240,0){
	\includegraphics[width=.4\columnwidth]{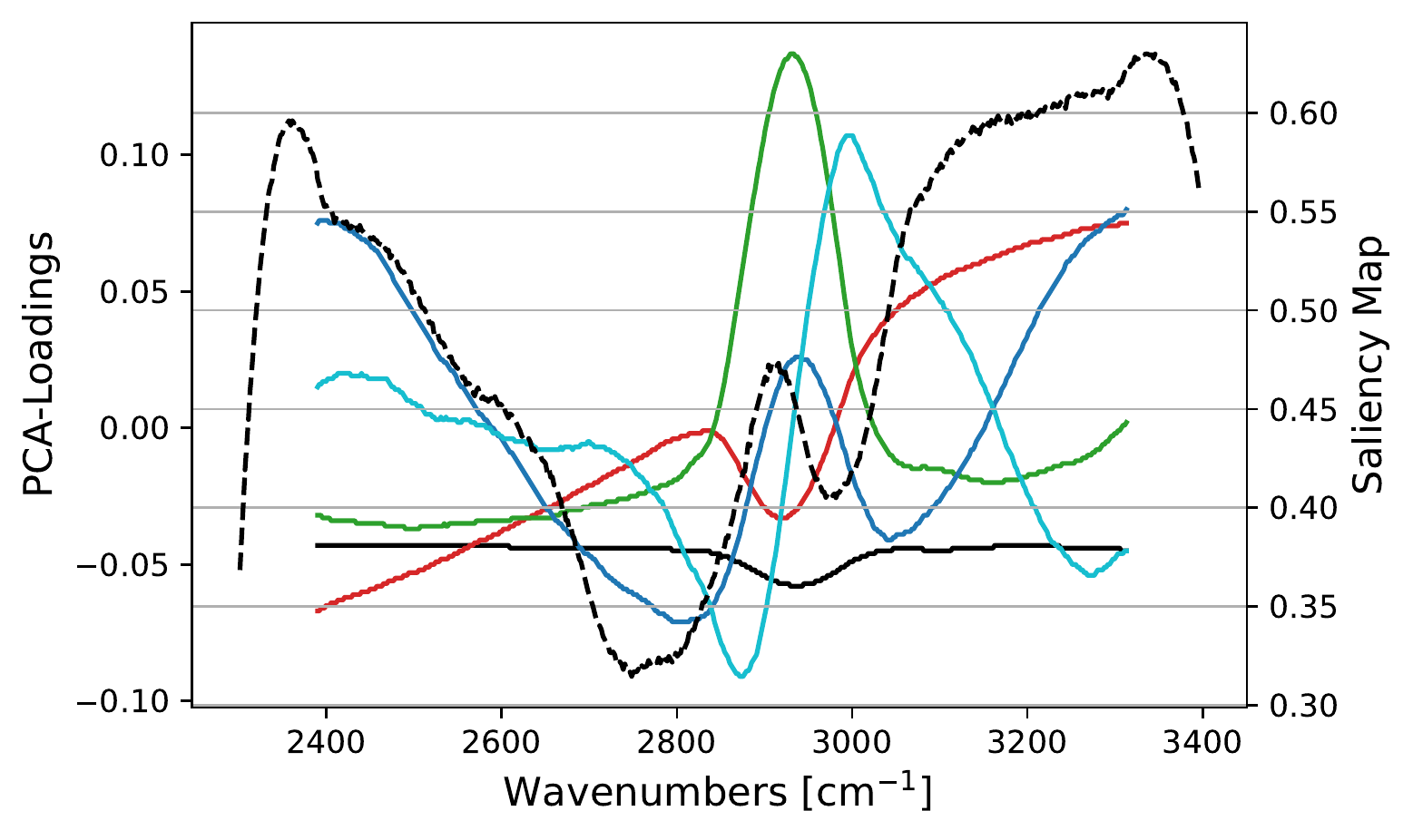}
        }
	\end{picture}
	\caption{SK--MEL--28 vs.~A375 comparison.
We report the PCA loadings using the same color code as in 
figure~\ref{f:fig03} and the salience map (dashed line) 
for the 1--D CNN applied to the spectra obtained by subtracting their 
mean value. 
}
\label{f:fig06}
\end{figure*}

In figure~\ref{f:fig03} we report the loadings 
computed for the first five PCA components, 
while Table~\ref{t:propvar} collects their values of standard deviation, 
proportion, and cumulative proportion of variance in both the HW and LW 
regions.  
Note that in both the regions the cumulative variance associated with these components exceeds the value $99,9\%$, with most of the variance related to the first principal component PC1. 
As shown in figure~\ref{f:fig03}, the contributions of the input variables to the first principal component -- the black line -- is almost 
constant in both the regions, with the exception for the peak 
at 234~cm$^{-1}$ (LW region) and 2930~cm$^{-1}$ (HW region). 
A peculiar behavior in correspondence of both the peaks is shown also by the second component, enforcing the hypothesis that these two peaks are the most relevant to achieve binary classification within this model. 


In figure~\ref{f:fig04} the average importance scores of the logistic 
regression on average pooling are shown.
As one can see, the LW region is particularly sensitive 
at 480--548~cm$^{-1}$ with importance scores 
of 0.48.
That region contains the peak associated to the silicon 
NWs (514~cm$^{-1}$, see figure~\ref{f:fig01}), 
whose classification role is likely associated to the different 
surface arrangement of DNA molecules on the SiNWs, 
resulting in a diverse substrate coverage and SERS enhancement factor, 
that in turn causes a different weight of the signal coming from 
the NWs with respect to the DNA one.
Instead, other main physical characteristics, still representing the 
interaction between the samples and the substrate, as the broad peak 
around 234~cm$^{-1}$ (see figure~\ref{f:fig01}) do not support the 
high performance of the logistic regression.
In the HW region, instead, one can see that the binary problem is
solved by exploiting information laying in sub--band
at 3200--3399~cm$^{-1}$.

In figure~\ref{f:fig05} the saliency maps of the 1D--CNN models are shown.
The saliency map presents two broad 
peaks around 310~cm$^{-1}$ and 490~cm$^{-1}$ with saliency values 
of 0.6 and 0.82 respectively.
Both these regions are not centered on some crucial 
wavenumbers such as 234~cm$^{-1}$ or 512~cm$^{-1}$; 
where physico--chemical spectral lines are usually located.
When dealing with the HW region, one can see that the saliency map 
reveals a flat region with saliency 0.6 at the right of the peak 
of the CH vibrations, i.e., 3030--3300~cm$^{-1}$.

In figure~\ref{f:fig06} we report the loadings related to 
the PCA components and the salience map for 
the 1--D CNN (left and right vertical axes respectively)
run on the spectra after having subtarcted their mean.
As far as LW is concerned, on the one hand 1--D CNN does not identify 
the region surrounding 200~cm$^{-1}$ as predictive, while the 
one after 300~cm$^{-1}$ is considered salient. On the other hand, 
the first PCA component detects that the spectrum intensity 
decreases until 300~cm$^{-1}$ to become then almost constant. 
Other PCAs, such as the third and the fourth components, provide 
a better representation of the characteristics at 230~cm$^{-1}$. 
It has to be remarked that the proportion of variance explained by 
these two components is 2 orders of magnitude smaller than the one 
associated to the first PCA component.

Regarding HW, the 1--D CNN model is capable to put in evidence 
the peak associated to the CH vibrations -- even with a salience 0.6, 
not particularly high. Within the PCA framework, also in this case the 
first components describes mainly a scaling factor, even if -- at 
least marginally -- also the band of the CH vibrations is represented. 
Higher components, for example the fourth and the fifth, put more in 
evidence the band located at about 2930~cm$^{-1}$.

\begin{table}
\begin{center}
\begin{tabular}{l|ccccc}
 & LRA & L2D & LRP & PCA & CNN \\
\hline
HaCaT vs.~CaCo--2 LW & 0.98 & 0.90 & 0.99 & 0.96 & 0.99 \\
HaCaT vs.~CaCo--2 HW & 0.97 & 0.73 & 0.99 & 0.98 & 0.99 \\
\hline
HaCaT vs.~HT29 LW & 0.75 & 0.90 & 0.93 & 0.99 & 0.93 \\
HaCaT vs.~HT29 HW & 0.97 & 0.98 & 0.99 & 1.00 & 0.94 \\
\hline
CaCo--2 vs.~HT29 LW & 0.94 & 0.87 & 0.99 & 1.00 & 0.97 \\
CaCo--2 vs.~HT29 HW & 0.83 & 0.93 & 0.93 & 0.99 & 0.99 \\
\hline
\end{tabular}
\end{center}
\caption{As in table~\ref{t:auc} for 
the three colon tumor 
cases listed at the beginning of Section~\ref{s:dataset}.
}
\label{t:co-auc}
\end{table}

\begin{figure*}[!htb]
	\begin{picture}(80,140)(-10,0)
        \put(0,0){
	\includegraphics[width=.4\columnwidth]{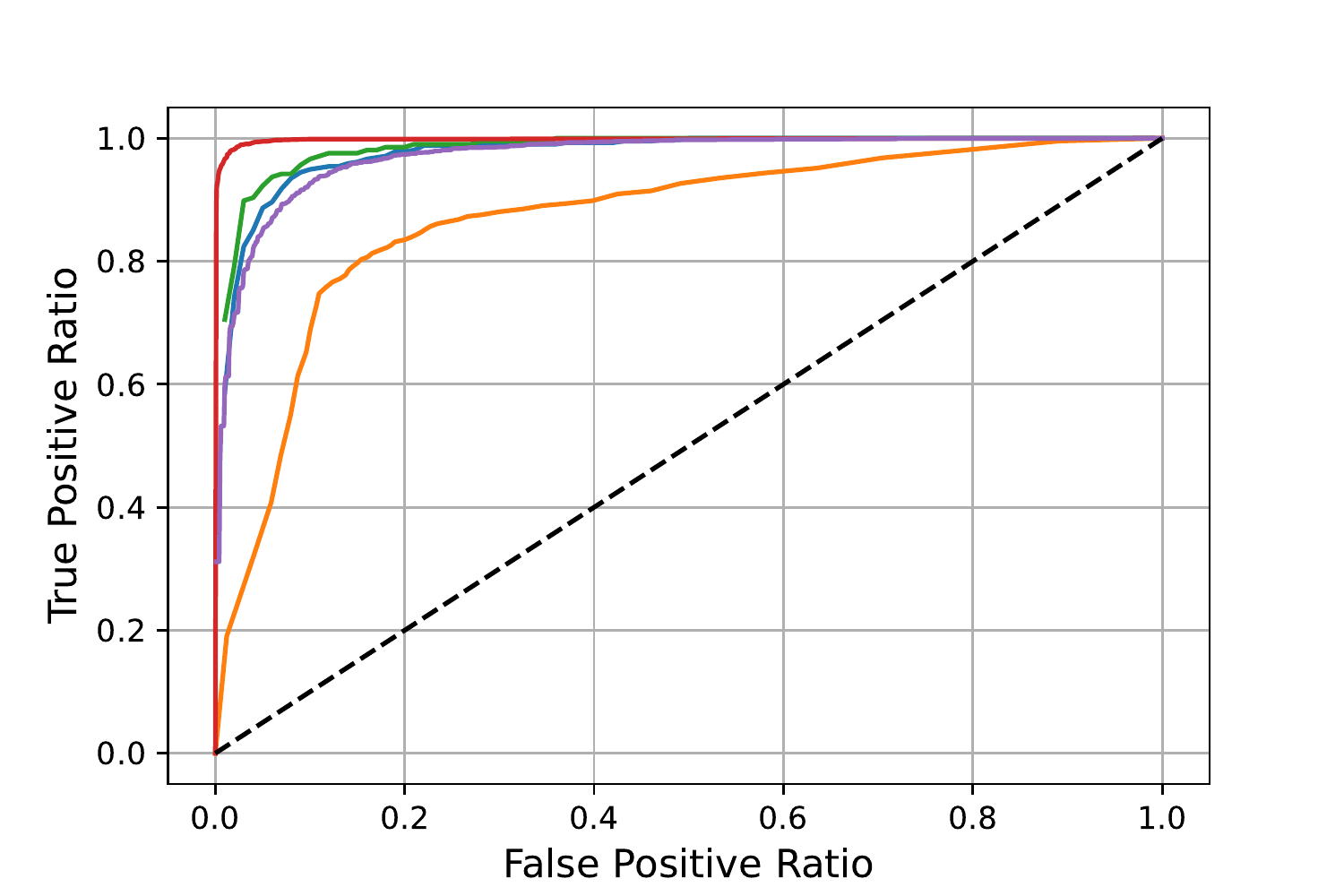}
        }
        \put(260,0){
	\includegraphics[width=.4\columnwidth]{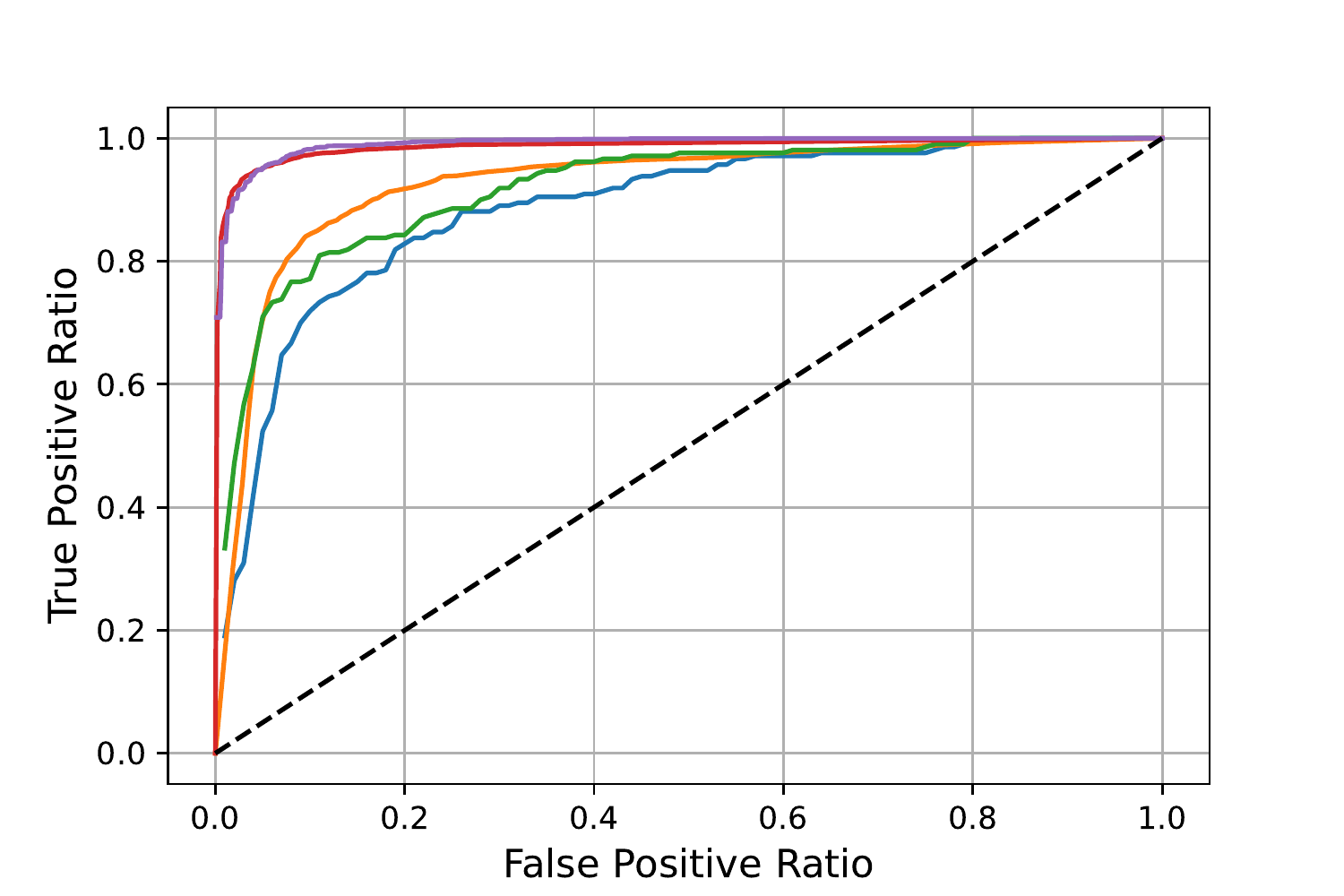}
        }
	\end{picture}
	\caption{As figure~\ref{f:fig02} for the case CaCo--2 vs.~HT29.}
\label{f:fig07}
\end{figure*}

\begin{figure*}[!htb]
	\begin{picture}(80,150)(-10,0)
        \put(0,0){
	\includegraphics[width=.4\columnwidth]{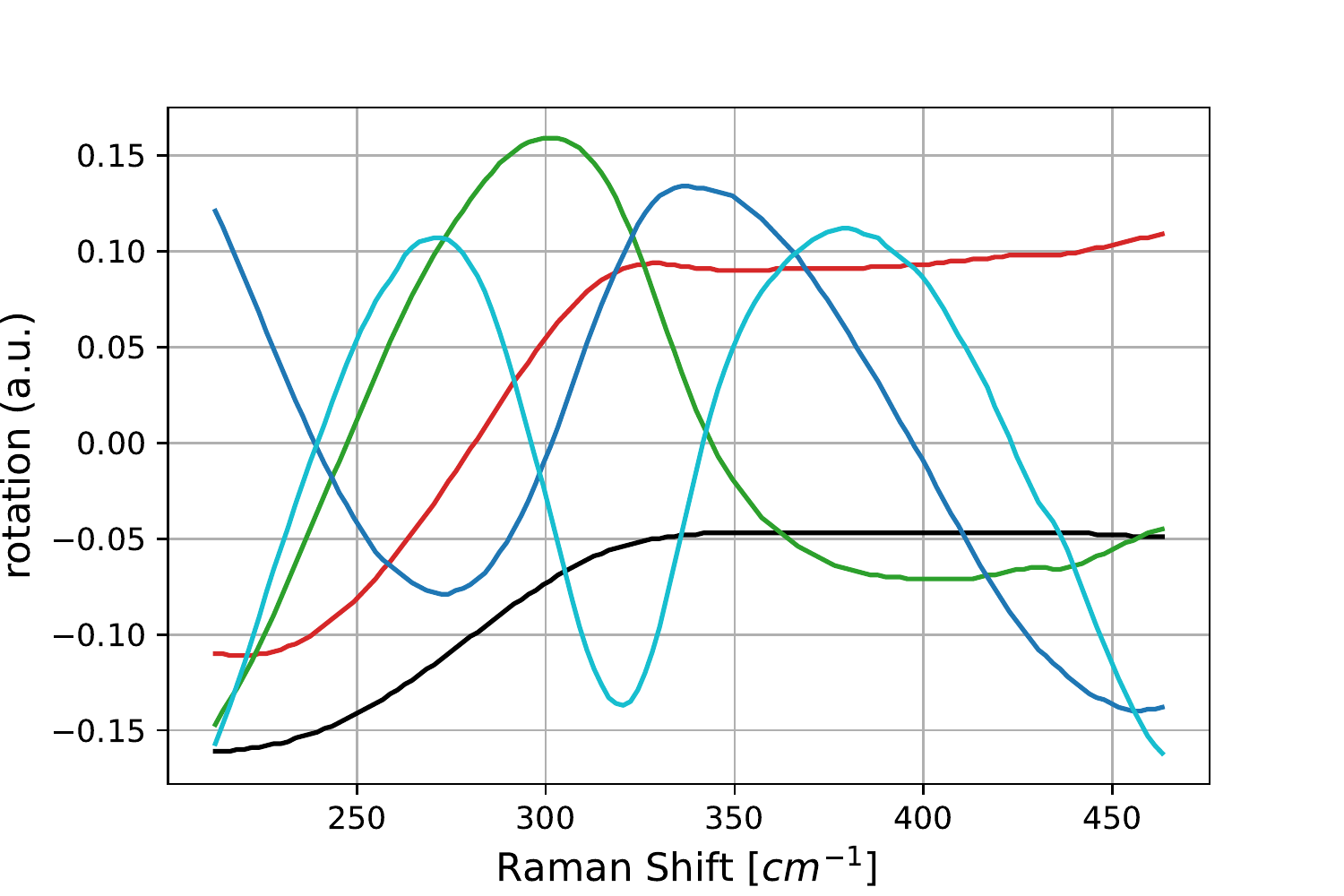}
        }
        \put(240,0){
	\includegraphics[width=.4\columnwidth]{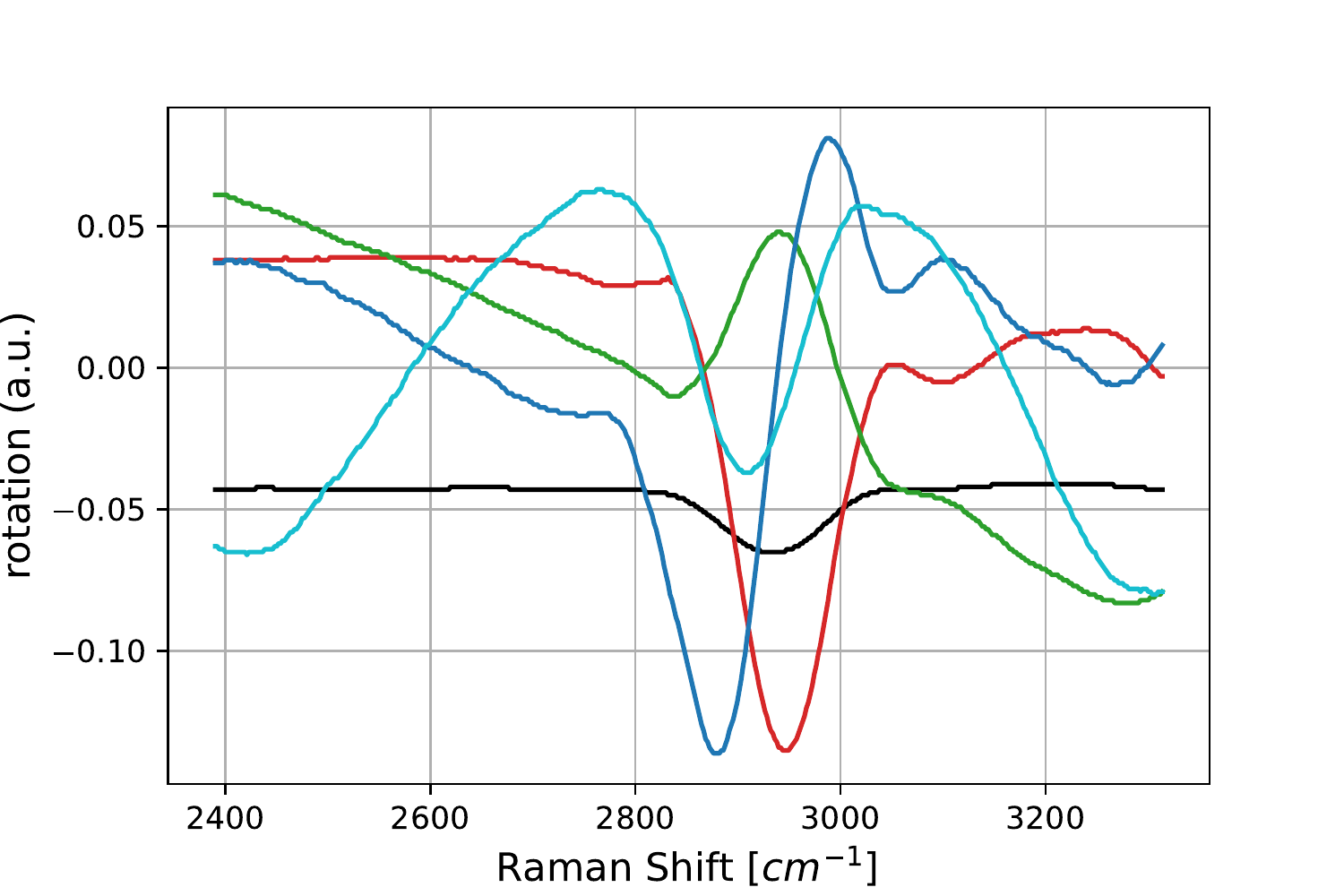}
        }
	\end{picture}
	\caption{As figure~\ref{f:fig03} for the case CaCo--2 vs.~HT29. Black, 
red, green, blue, and light blue identify principal components from 
one to 5 respectively.}
\label{f:fig08}
\end{figure*}

\subsection{Analysis of colon tumor data}
\label{s:colon} 
\par\noindent 
As explained above, we tested the robustness of the proposed methods 
by applying our analysis to a data set obtained by using 
different tumor samples, namely, the colon tumor cells 
listed as cases 4--6 in Section~\ref{s:dataset}.
Although the average spectra compare each other differently with 
respect to the melanoma case, see figure~\ref{f:fig01}, we will show 
that our techniques perform well also in this case. 

We report in Table~\ref{t:co-auc} the AUC values obtained with the 
different methods for the high and low wavenumber regions of the 
spectra.
The associated ROC graphs are reported in 
figure~\ref{f:fig07}.

We note that, although the global methods do not perform perfectly 
in some of the cases, the local ones achieve absolutely 
remarkable performances in classifying tumor vs.~healthy samples
and also in classifying 
the two different tumor phenotypes (CaCo--2 vs.~HT29 
comparison). 

In the remaining part of this section 
we restrict our discussion to the CaCo--2 vs.~HT29
comparison and show how our statistical analysis provides
information at the physical and chemical level.

\begin{figure*}[!htb]
	\begin{picture}(80,140)(-10,0)
        \put(0,0){
	\includegraphics[width=.4\columnwidth]{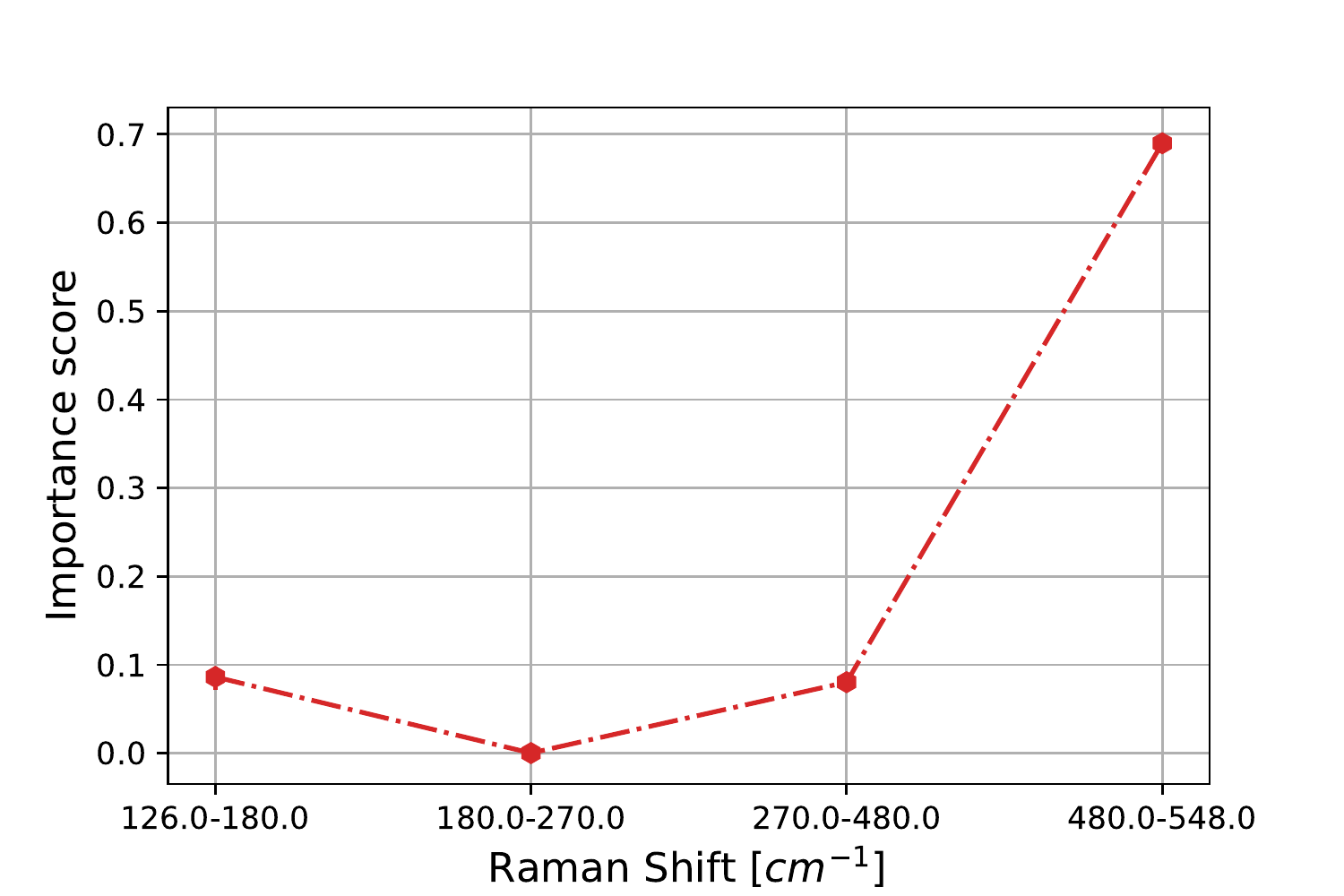}
        }
        \put(240,0){
	\includegraphics[width=.4\columnwidth]{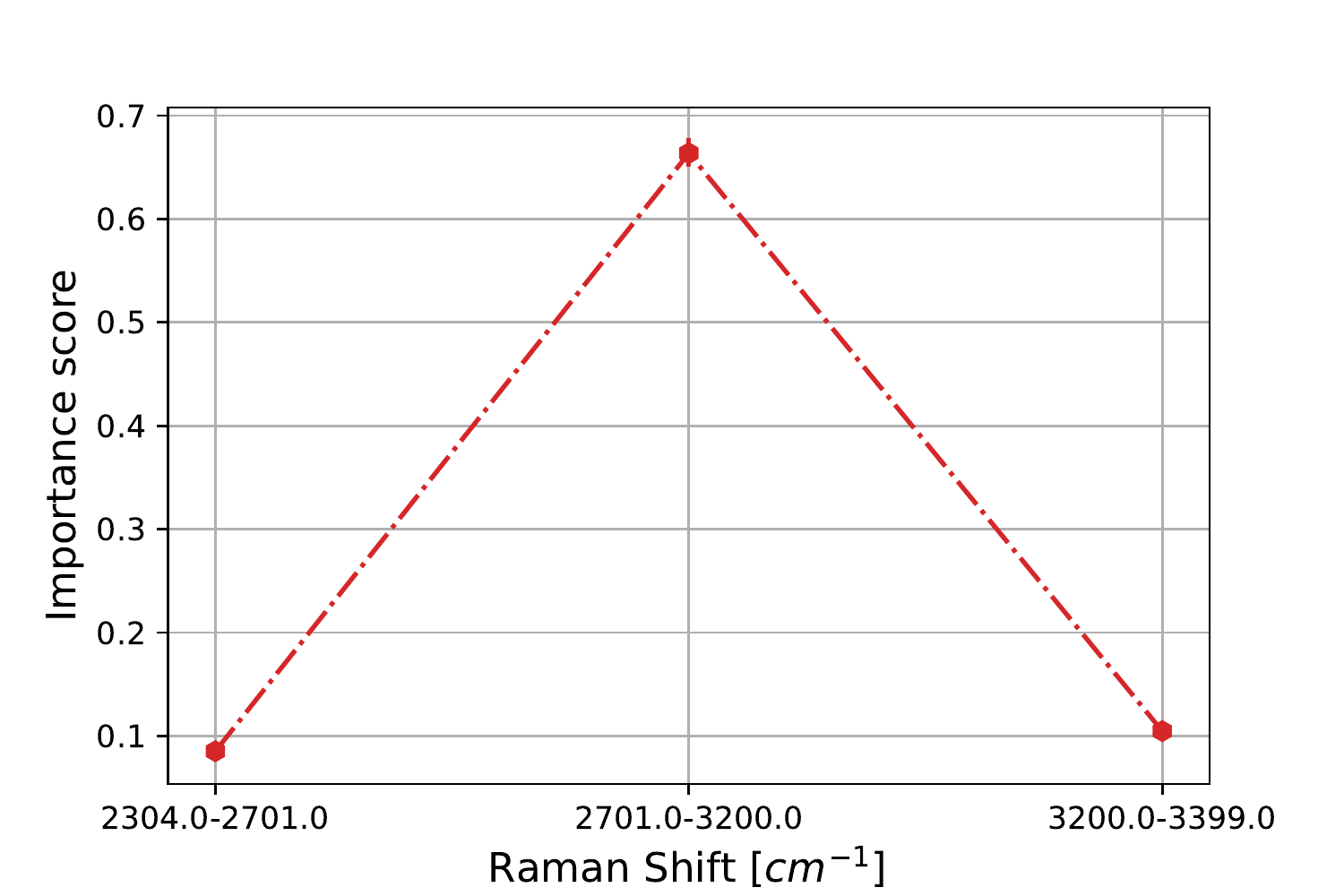}
        }
	\end{picture}
	\caption{As figure~\ref{f:fig04} for the case CaCo--2 vs.~HT29.}
\label{f:fig09}
\end{figure*}

\begin{figure*}[!htb]
	\begin{picture}(80,140)(-10,0)
        \put(0,0){
	\includegraphics[width=.4\columnwidth]{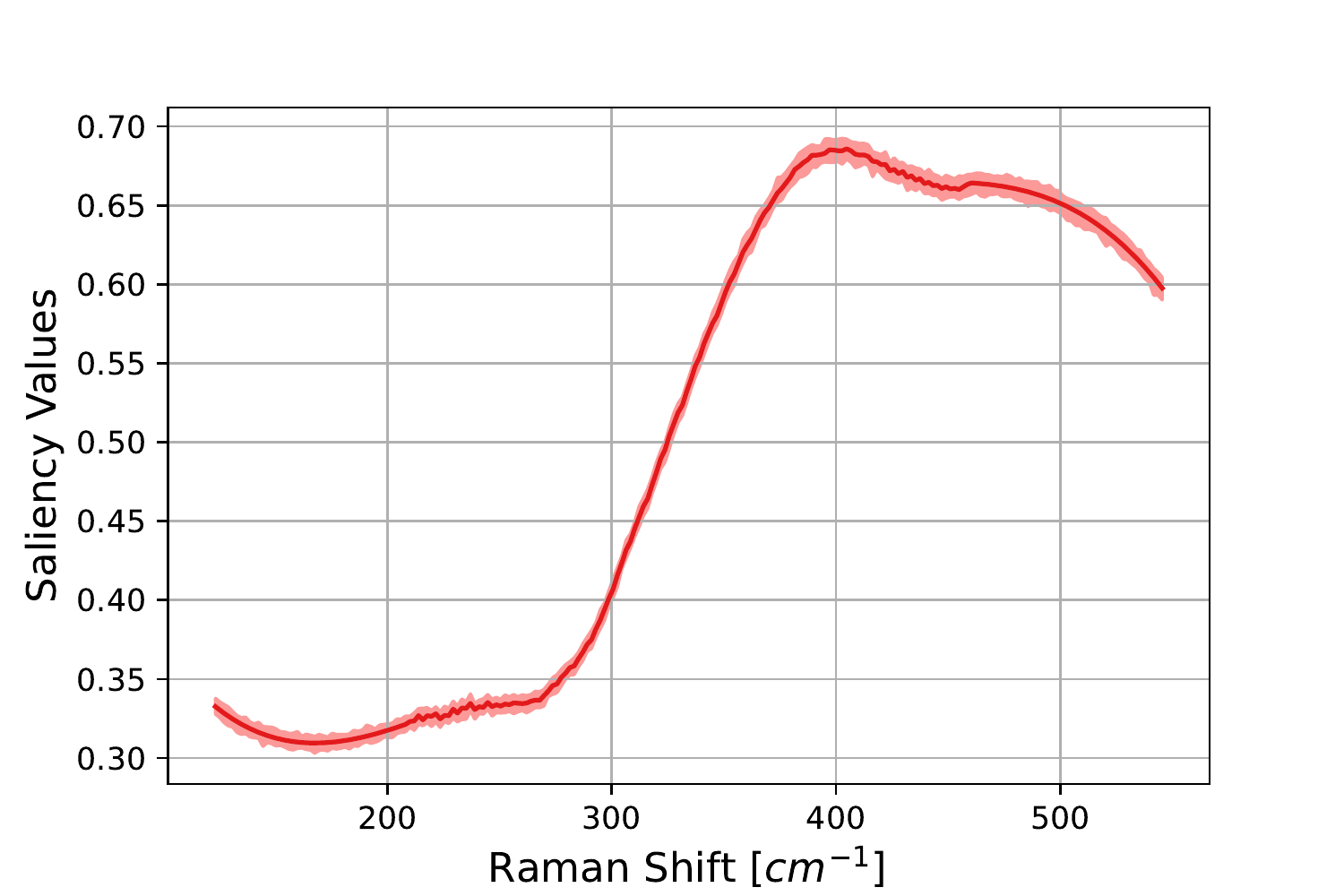}
        }
        \put(240,0){
	\includegraphics[width=.4\columnwidth]{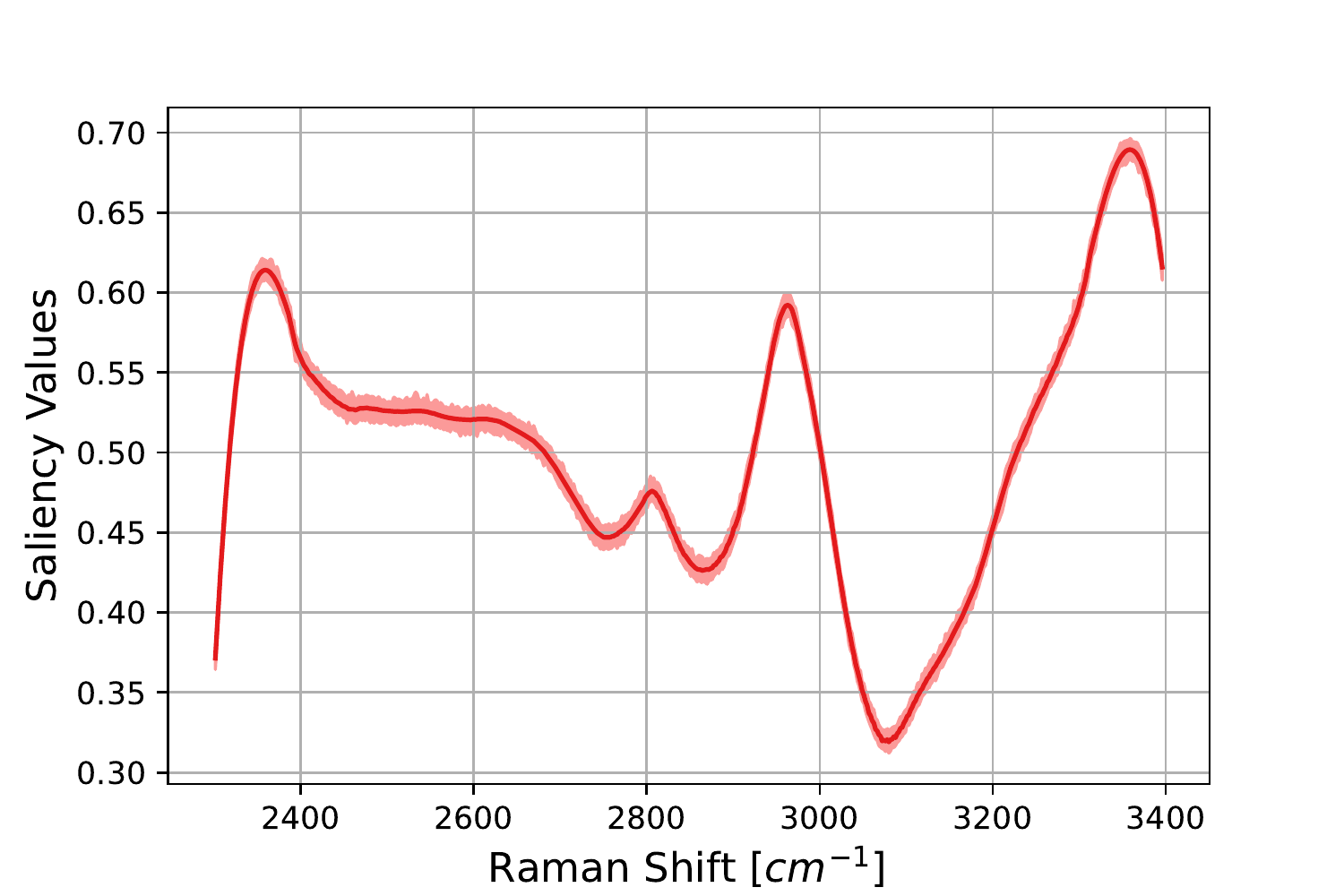}
        }
	\end{picture}
	\caption{As figure~\ref{f:fig05} for the case CaCo--2 vs.~HT29.}
\label{f:fig10}
\end{figure*}

In figure~\ref{f:fig08} we report the loadings 
computed for the first five PCA components.
Regarding the LW region (left panel), the contributions of 
the input variables  to the first principal component -- the black line -- 
are larger at lower frequencies, to became then constant at higher ones. 
In the same region, the second principal component is characterized by 
similar contributions, but opposite in sign. Finally, in the third 
component a peak is showed at about $310$~cm$^{-1}$.
In the HW region (right panel), the contributions are almost constant 
for all the involved frequencies, with the exception of a small 
peak at $2930$~cm$^{-1}$. Other higher peaks characterize in the same 
region also the second and third principal components, suggesting its 
importance to achieve binary classification. 

In figure~\ref{f:fig09} the average importance scores of the logistic 
regression on average pooling are shown.
As for the melanoma case, 
the LW region is particularly sensitive 
at 480--548~cm$^{-1}$ with larger importance scores 
of order 0.69.
Similary to the SK--MEl--28 vs.~A375 case, 
the broad peak 
around 234~cm$^{-1}$ (see figure~\ref{f:fig01}) does not support the 
high performance of the logistic regression.
In the HW region, instead, the binary problem is 
solved by exploiting information laying in different sub--bands
with respect to the ones relevant in th melanoma case.
More specifically, 
a high importance score 
of 0.65 is found in the sub--band 2701--3200~cm$^{-1}$, i.e., where is 
located the broad peak representing the vibrational modes
of CH--groups (usually represented by a broad peak centered 
at 2930~cm$^{-1}$; see, figure~\ref{f:fig01}).

In figure~\ref{f:fig10} the saliency maps of the 1D--CNN models are shown.
As one can see, the cases SK--MEL--28 vs.~A375
and CaCo--2 vs.~HT29 support in a different way the highly accurate 
predictions of the CNN model.
As already noted above, 
the saliency map of case SK--MEL--28 vs.~A375 presents two broad 
peaks around 310~cm$^{-1}$ and 490~cm$^{-1}$.
Likewise, the saliency map of case CaCo--2 vs.~HT29
reveal a salient region at 350--514~cm$^{-1}$.
When dealing with the HW region, one can see that the saliency map 
of case CaCo--2 vs.~HT29 can reveal a narrow salient region in correspondence 
of 2930~cm$^{-1}$ (a peak with saliency value 0.6).

\section{Conclusions}
\label{s:conclusioni} 
\par\noindent
In this paper we have demonstrated via a thorough statistical 
analysis that Raman mapping obtained by dropping genomic DNA on 
disordered Ag/SiNWs can be used to classify different subtypes of 
tumors both for melanoma and colon cancer. 

We have developed several statistical approaches to classify the 
experimental data showing that both the global and local methods are 
able to distinguish the healthy and malignant DNA molecules through a 
different interaction of these molecules with the Ag/SiNWs platform,
mainly affecting the low wavenumber region of the analysed spectral range.
In addition, the local methods achieve absolutely remarkable performances 
in classifying the diverse tumor phenotypes. Indeed, 1--D CNN model 
takes advantage of its pattern recognition activity to capture the 
characteristic bands associated with the CH vibrations,
located in the high wavenumber region of the analysed spectral region,
by allowing 
to separate different phenotypes for both melanoma and colon cancer 
on the basis of their methylation degree. 

A similar result is achieved by means of the method based on a PCA 
decomposition, which exploits subsets of relevant frequencies to 
perform a logistic regression. We highlight that the discrimination 
of DNA coming from distinct malignant phenotypes occurs without any 
knowledge of the basis sequencing, unlike most of the present biochemical 
methods. Our results thus suggest that the analysis of Raman spectra of 
genomic DNA directly dropped onto disordered Ag/SiNWs with 1--D CNN 
and PCA algorithms provides a rapid, simple, and accurate discrimination 
of the cancer subtypes by offering a powerful and effective guidance in 
the patient personalized treatment.

\begin{acknowledgments}
The research activity is funded by Regione Lazio within the project DIANA,
``DIAgnostic potential of disorder: development of an innovative 
NAnostructured platform for rapid, label--free and low--cost analysis 
of genomic DNA'', POR FESR Lazio 2014 -- 2020, 
Progetti Gruppi di ricerca call 2020, 
A0375--2020--36589, CUP B85F21001240002.

A.C. acknowledges the support of the Italian Minister of Foreign Affairs and
International Collaboration
(MAECI) under the Joint research project ``Scalable nano--plasmonic platform 
for differentiation
and drug response monitoring of organtropic metastatic cancer cells''
(US19GR07).
\end{acknowledgments}



\vfill\eject

\bibliographystyle{unsrt}

\end{document}